\documentclass[epj]{svjour}
\usepackage{latexsym}
\usepackage{psfig}

\begin{document}
\title{Photoemission, inverse photoemission and superconducting 
correlations in Hubbard and t--J ladders:
role of the anisotropy  between legs and rungs} 
\author{
J.~Riera\inst{1,2}
\and D.~Poilblanc\inst{1}
\and E.~Dagotto\inst{3}
}
\institute{Laboratoire de Physique Quantique \& Unit\'e mixte
de Recherche CNRS 5626,
Universit\'e Paul Sabatier, 31062 Toulouse, France
\and
Instituto de F\'{\i}sica Rosario, Consejo Nacional de 
Investigaciones 
Cient\'{\i}ficas y T\'ecnicas y Departamento de F\'{\i}sica,
Universidad Nacional de Rosario, Avenida Pellegrini 250, 2000-Rosario,
Argentina
\and
Department of Physics and National High Magnetic Field Lab,
Florida State University,
Tallahassee, FL 32306, USA
}
\date{Received: April 20}

\abstract{
Several experiments in the context of ladder materials have
recently shown that the
study of simple models of anisotropic ladders (i.e. with different couplings along legs and
rungs) is important for the understanding of these compounds. In this
paper Exact Diagonalization studies of the one-band Hubbard and $t-J$
models are reported for a variety of densities, couplings, and
anisotropy ratios. The emphasis is given to the one-particle spectral
function $A({\bf q},\omega)$
which presents a flat quasiparticle dispersion at the chemical potential in some
region of parameter space. This is correlated with the existence of
strong pairing fluctuations, which themselves are correlated with 
an enhancement of the bulk-extrapolated value for the two-hole
binding energy  as well as with the strength of 
the spin-gap in the hole-doped system. 
Part of the results for the spectral function are explained using a
simple analytical picture valid when the hopping along the legs is small.
In particular, this picture predicts an insulating state at
quarter filling in agreement with the metal-insulator transition observed at
this special filling for increasing rung couplings. 
The results are compared against previous literature, and in addition
pair-pair correlations using extended operators are also here reported.
} 
\titlerunning{Photoemission, inverse photoemission and superconducting 
correlations in Hubbard and t--J ladders}
\maketitle

\section{Introduction}

The discovery of superconductivity in quasi-two dimensional (2D)
copper-oxide materials has led to a renewed interest in the physics
of doped Mott-Hubbard insulators and in the interplay between
magnetism and superconductivity.
Recently another class of copper oxide materials based on weakly coupled
one dimensional (1D) ladders~\cite{Review_ladder}
has been synthetised.
Their structure is closely related to the one of the 
2D perovskites, namely it contains
S=1/2 copper spins which are antiferromagnetically coupled
along the ladder direction (legs) and along the rungs through
Cu-O-Cu bonds.
Recent experimental results reporting  
superconductivity~\cite{Uehara96} in the 
hole-doped ladder cuprate
Sr$_{14-x}$Ca$_x$Cu$_{24}$O$_{41}$ have clearly established that
the existence of a superconducting state is not restricted to 2D doped
antiferromagnets but it actually extends to a wider class of strongly
correlated copper-oxide materials. 
Thus, the novel ladder compounds offer new perspectives, both for experimentalists
and theorists, to understand the mechanism of superconductivity 
in strongly correlated low-dimensional systems. 

Ladder structures can also be found in other oxides such as 
vanadates~\cite{Millet}.
Magnetic susceptibility measurements on MgV$_2$O$_5$ have been interpreted 
in terms of weakly coupled Heisenberg ladders. In addition, recent X-rays 
scattering experiments~\cite{xrays_NaV2O5} have suggested that NaV$_2$O$_5$
could be considered as a quarter-filled ladder system. 

While the stoichiometric parent
compounds of the superconducting 2D cuprates are 
antiferromagnetic Mott insulators, the parent
insulating ladders exhibit spin liquid properties.
The existence of a spin gap in a spin-ladder
structure has been first proposed
theoretically~\cite{Dagotto92}
and found experimentally in several even-leg ladder copper-oxide systems
(such as SrCu$_2$O$_3$~\cite{Azuma94,Review_ladder} and
 LaCuO$_{2.5}$~\cite{Hiroi95}).

It has been suggested that the spin gap, if robust under doping,
could be responsible for an attractive
interaction between holes on the same rung~\cite{Dagotto92,pairing}. 
Although recent experiments~\cite{Jerome} suggest that the spin gap 
disappears in hole-doped ladders at the high pressure needed to achieve
superconductivity, part of the
spin excitations are still suppressed as the temperature is lowered
in the normal state
as predicted theoretically~\cite{modes_ladder}.
Such a behavior bears similarities with the pseudogap behavior of  the 
underdoped 2D superconducting cuprates. 

It is expected that the formation of hole pairs on the rungs can 
lead to competing superconducting pairing
or 4k$_F$ Charge Density Wave (CDW) correlations as 
e.g. found in the weak coupling limit~\cite{weak_coupling}.
For isotropic $t-J$ ladders (i.e. with equal couplings along legs and
rungs), it has been established that 
the d-wave-like superconducting pairing correlations~\cite{Hayward95} 
are dominant in a large region of the phase diagram~\cite{phase_diagram}. 
Such a state, which is referred to as C1S0 in the language of
the RG analysis~\cite{noteCnSm}, is characterized by a single
gapless charge mode and a 
gap in the spin excitation spectrum~\cite{modes_ladder}
and belongs to the same universality class as the Luther-Emery phase
of the 1D chain.
Because of the one-dimensional character 
of the system, the superconducting correlations at large distances 
still behave following power laws~\cite{weak_coupling,phase_diagram}.
Using finite size scaling analysis and conformal invariance 
relations~\cite{phase_diagram}, the corresponding critical exponents have been
computed in the case of equal rung and leg couplings (isotropic case).
Note that a small Josephson tunneling between the ladders
is expected to give a finite superconducting critical 
temperature~\cite{Orignac}. At finite doping, 
the spin gap is expected to vanish below a small critical $J/t$ 
ratio~\cite{phase_diagram,mueller}. Presumably, such a transition is 
associated to an instability of the hole pairs gas towards a liquid made
out of
individual holes with two spin and two charge collective modes (C2S2 
phase)~\cite{mueller}. Whether the disappearance of the spin gap
observed by NMR experiments in the doped Sr$_{14}$Cu$_{24}$O$_{41}$ 
superconducting ladder material under pressure~\cite{Jerome} is
connected to such a transition is under much debate. 

Although the isotropic case is the most widely analyzed situation in the
context of theories for ladders,
it has now become clear that most of the actual ladder materials have in fact
different leg ($J_\parallel$) and rung ($J_\perp$) magnetic couplings and/or 
hopping matrix elements $t_\parallel$ (legs) and $t_\perp$ (rungs). 
Recent neutron scattering 
experiments~\cite{Eccleston} on
the insulating ladder Sr$_{14}$Cu$_{24}$O$_{41}$ actually suggest a ratio of 
$J_\perp/J_\parallel\simeq 0.5$. A similar anisotropy was in fact
also predicted previously in the context of the 
SrCu$_{2}$O$_{3}$ material~\cite{Johnston}.
On the other hand, the vanadate ladder
NaV$_2$O$_5$ apparently corresponds to the opposite limit of strong rung couplings
with a ratio $t_\perp/t_\parallel\simeq 2$~\cite{Horsch} which could
justify a description at quarter-filling in terms of an effective 1D Heisenberg 
model~\cite{Horsch,Augier1}. 

The purpose of the present paper is to investigate spectral properties
and superconducting correlations of anisotropic 
Hubbard and $t-J$ ladders by exact diagonalization methods. 
Finite size scaling analysis is used to obtain 
physical quantities such as the pair binding energy or the spin gap.
Dynamical correlations functions are also computed using a continued
fraction method. 
The focus of the paper will be on the role of the ladder anisotropy
(regulated by the ratios of the rung couplings to the leg couplings) as
well as on the influence of doping.
The anisotropic Hubbard ladder is defined as,
\begin{eqnarray}
H&=&t_\parallel \sum_{i,\alpha,\sigma} (c_{i,\alpha;\sigma}^\dagger
c_{i+1,\alpha;\sigma} + h.c.)\nonumber \\
&+&t_\perp \sum_{i,\sigma} (c_{i,1;\sigma}^\dagger c_{i,2;\sigma}+h.c.)
\label{Hubbard}\\
&+&U\sum_{i,\alpha} n_{i,\alpha;\uparrow} n_{i,\alpha;\downarrow}\, ,
\nonumber
\end{eqnarray}
where the index $\alpha$ stands for the chain index ($=1,2$).
Anisotropy ratios $r_a=t_\perp/t_\parallel$ in the range $0.5\le r_a\le 2.5$
will be considered.
Most of the calculations reported below have been carried out in the strong 
coupling regime $U/t_\parallel=8$.
Motivated by the doped cuprate and vanadate ladders,
the studies below are performed in the electron density range
$0.5\le n\le 1$.
For $U/t_\parallel\gg 1$ and $U/t_\perp\gg 1$, the low energy 
spin and charge degrees of freedom can be described by the effective 
anisotropic $t-J$ ladder with doubly occupied sites projected out,
\begin{eqnarray}
H&=&J_\parallel \sum_{i,\alpha} ({\bf S}_{i,\alpha}\cdot
{\bf S}_{i+1,\alpha}-\frac{1}{4}n_{i,\alpha}n_{i+1,\alpha})
\nonumber \\
&+&J_\perp \sum_{i} ({\bf S}_{i,1}\cdot {\bf S}_{i,2}
-\frac{1}{4}n_{i,1}n_{i,2}) \label{tJ} \\
&+&t_\parallel \sum_{i,\alpha,\sigma}
({\tilde c}_{i,\alpha;\sigma}^\dagger{\tilde c}_{i+1,\alpha;\sigma}
+ h.c.)\nonumber \\
&+&t_\perp \sum_{i,\sigma} ({\tilde c}_{i,1;\sigma}^\dagger
{\tilde c}_{i,2;\sigma}+h.c.) \,   ,
\nonumber
\end{eqnarray}
where 
${\tilde c}_{i,\alpha;\sigma}^\dagger
= c_{i,\alpha;-\sigma}(1-n_{i,\alpha;\sigma})$ 
are {\it hole} Guzwiller projected creation
operators. The large-$U$ limit of the
anisotropic Hubbard ladder leads to antiferromagnetic exchange
couplings of the form $J_\beta=4t_\beta^2/U$ in the two directions
$\beta=\parallel,\perp$. 
Hence, for simplicity, the relation $J_\perp/J_\parallel
=(t_\perp/t_\parallel)^2$ will be here assumed even outside of the range 
$J_\perp/t_\perp\ll 1$ and $J_\parallel/t_\parallel\ll 1$ of rigorous validity
of the equivalence between the two models.
In the rest of the paper, energies will be measured in units 
of $t_\parallel$ unless specified otherwise.

\section{Single particle spectral function}

\subsection{Motivations}

Let us examine first the one-particle spectral function, 
\begin{equation}
A({\bf q},\omega)=A_e ({\bf q},\omega) + A_h ({\bf q},\omega) \,  ,
\end{equation}
where $A_e ({\bf q},\omega)$ corresponds to the 
density of the unoccupied electronic states,
\begin{equation}
A_e({\bf q},\omega)=-\frac{1}{\pi}{\rm Im} 
\big< c_{{\bf q};\sigma}
\frac{1}{\omega+i\epsilon-H+E_0}
c_{{\bf q};\sigma}^\dagger\big>_0 \, ,
\end{equation}
and $A_h ({\bf q},\omega)$ corresponds to the 
density of the occupied electronic states (ie unoccupied hole states),
\begin{equation}
A_h({\bf q},\omega)=-\frac{1}{\pi}{\rm Im} 
\big< c_{{\bf q};\sigma}^\dagger
\frac{1}{\omega+i\epsilon+H-E_0}
c_{{\bf q};\sigma}\big>_0 \, .
\end{equation}
Here $\big<...\big>_0$ stands for the expectation value in the ground state
wave function of energy $E_0$. The transverse component of the
momentum ${\bf q}$ takes only the two values $q_\perp=0,\pi$ corresponding 
to symmetric or anti-symmetric states with respect to the reflection
exchanging the two chains. 
$A_h({\bf q},\omega)$ and $A_e({\bf q},\omega)$ are of crutial 
importance since they can be directly measured in angular-resolved 
photoemission (ARPES) and inverse photoemission (IPES) spectroscopy experiments,
respectively.

Thus far, the role of the anisotropy ratio $t_\perp/t_\parallel$ in
dynamical properties of ladders has not been studied in detail,
 except at half-filling $n=1$. In this case,
the spin gap is remarkably robust and persists down to
arbitrary small interchain magnetic coupling $J_\perp$~\cite{SG_anisotropy}.
The single particle (and two particles) spectral functions of the
Hubbard ladder have been obtained at $n=1$ using quantum Monte Carlo (QMC)
simulations~\cite{Hubbard_neq1}. Working at $U =8$, two regimes were
identified~\cite{Hubbard_neq1} depending on the magnitude of the ratio 
$t_\perp/t_\parallel$. For instance, increasing $t_\perp$ the half-filled
Hubbard ladder evolves from a four-band (at small $t_\perp/t_\parallel$) to
a two-band (at large $t_\perp/t_\parallel$) insulator. The latter regime can 
be understood from the non-interacting $U\sim 0$ picture: in this case, 
two heavily weighted bonding ($q_\perp=0$)
and anti-bonding ($q_\perp=\pi$) bands are separated by
an energy $\sim 2t_\perp$ and the Fermi level lies in between. 
On the other hand, a small fraction of the total spectral weight 
is found in the inverse photoemission 
spectrum ($\omega >\mu$) for $q_\perp=0$ and in the
photoemission spectrum ($\omega <\mu$) for $q_\perp=\pi$.
In fact, in the $t_\perp/t_\parallel > 1$ limit, the spin-spin correlation 
length is very short~\cite{Hubbard_neq1} and a description in terms
of a rung Hamiltonian (reviewed in the next section) is appropriate 
($t_\parallel$ can then be treated as a small perturbation).

In the other limit $t_\perp/t_\parallel < 1$ of two weakly coupled chains
the magnetic correlation length along the chains direction becomes larger.
Although no magnetic long range order exists, a description of the 
single particle properties in terms of a Hartree Fock spin-density-wave
(SDW) picture turns out to be reasonably accurate~\cite{Hubbard_neq1}. 
For both $q_\perp=0$ and
$q_\perp=\pi$ a dispersive structure is observed with a (SDW-like) 
gap $\sim U$ separating the photoemission and inverse photoemission 
energy regions.
It is worth noticing that the low energy electron or hole excited states
now occurs at momentum ${\bf q}=\pi/2$ in contrast to the large  
$t_\perp/t_\parallel$ limit where they occur at momenta ${\bf q}=\pi$ 
($\omega <\mu$) and ${\bf q}=0$ ($\omega >\mu$).
The Hartree-Fock treatment correctly predicts a bandwidth of 
order $J_\parallel$ due to the magnetic scattering (similar to the
spinon-like excitations of the single chain~\cite{Maekawa}). 
However, it fails to reproduce the broad incoherent background 
reminiscent of the holon excitations of the single chain~\cite{Maekawa}.

Away from half-filling (n=1) QMC simulations face 
the well known "minus sign" problem (especially at low temperature and
large $U$) which increases the statistical 
errors and, hence, 
reduces the accuracy of the analytic continuation to the real frequency
axis. Thus far, QMC studies of the doped Hubbard ladder have been restricted
to $U\le 4$ in the range 
$1.4\le t_\perp/t_\parallel\le 2$ and for temperatures 
larger than $t_\parallel/8$~\cite{Noack}. Density matrix renormalization
group techniques, on the other hand, can currently only provide information
about static correlations~\cite{Noack}.
A recently developed variational technique based
on the use of a reduced Hilbert space once the ladder problem is
expressed in the rung-basis~\cite{new1} can produce accurate dynamical
results on $2 \times 20$ 
clusters at zero temperature and finite hole density~\cite{new2}. 
However, this technique has been applied only to isotropric ladders 
thus far.
In the present work, alternative approaches have been used.
First, following Ref.~\cite{Hubbard_neq1},
a simple estimation of $A({\bf q},\omega)$ 
in the single rung approximation has been carried out. This calculation is
valid in the limit of small 
$t_\parallel$ and it is useful in order to discuss the possible 
existence of metal-insulator transition at commensurate densities such as
$n=0.5$ or $n=0.75$. This simple analytical 
scheme gives also a basis for understanding
more elaborate numerical calculations.
In a second step, exact diagonalization studies based
on the Lanczos algorithm were performed to investigate a broad region of
 parameter space.

\subsection{Local rung approximation: metal-insulator transition}

\begin{figure}[htbp]
\begin{center}
\psfig{figure=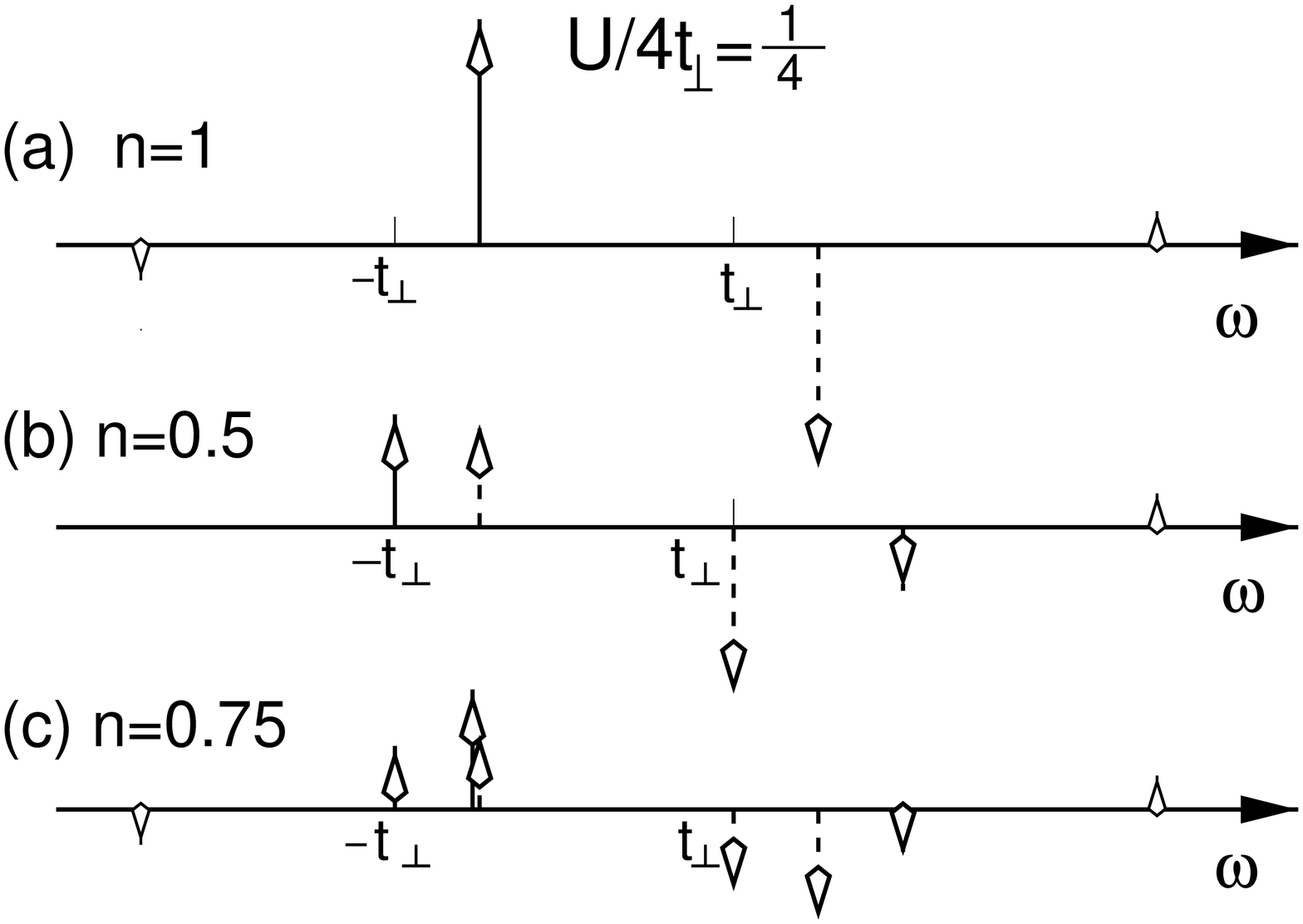,width=8truecm,angle=0}
\psfig{figure=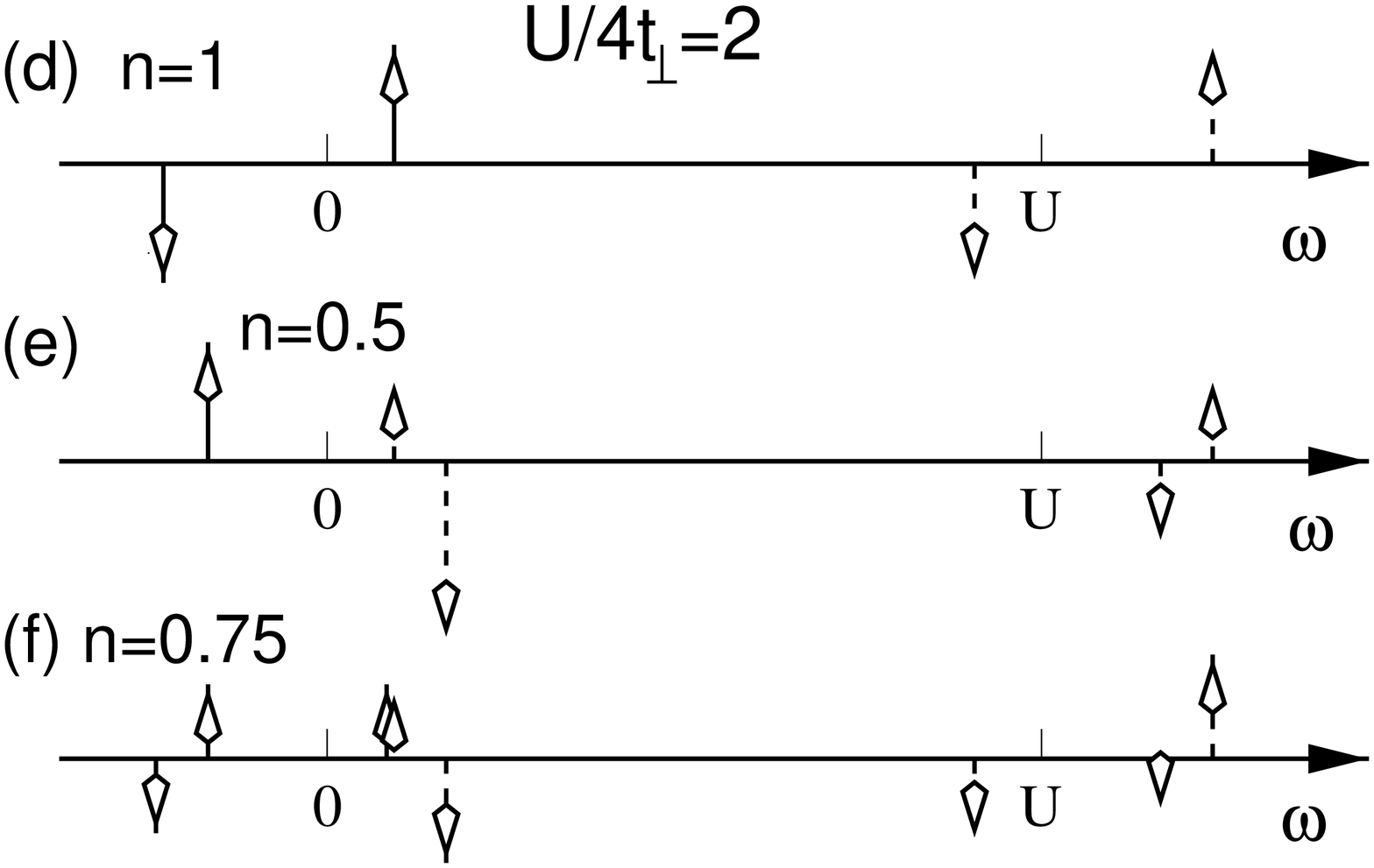,width=8truecm,angle=0}
\end{center}
\caption{
Schematic representation of the single particle 
spectral function vs frequency, in the $t_\parallel\rightarrow 0$ limit.
The position of the single particle energy poles are indicated by arrows
whose lengths are proportional to the spectral weights associated to
the corresponding transitions. Arrows pointing upwards (downwards) correspond 
to $q_\perp=0$ ($q_\perp=\pi$). 
The photoemission peaks (occupied states) and the inverse photoemission peaks
(empty states) correspond to full line and dashed line arrows, respectively.
The spectra are shown for ratios 
$\frac{U}{4t_\perp}\simeq 1/4$ ((a), (b) and (c)) 
and $\frac{U}{4t_\perp}\simeq 2$ ((d), (e) and (f))
and for electron densities $n=1$, $n=0.5$ and $n=0.75$ as 
indicated.
}
\label{Akw_rung}
\end{figure}

Let us consider the limit where $t_\parallel$ is the smallest 
energy scale, i.e.  $t_\parallel\ll t_\perp$ and $t_\parallel\ll U$.
First,  $A(q_\perp,\omega)$ can be calculated straightforwardly at
densities $n=1$ and $n=0.5$ by diagonalizing exactly the single rung 
Hamiltonian for 0, 1, 2 and 3 particles 
(see Ref.~\cite{Hubbard_neq1} for details). 
At half-filling one obtains,
\begin{eqnarray}
A(0,\omega)&=&\alpha^2 \, \delta (\omega-\Omega(2,1)) + 
(1-\alpha^2)\, \delta (\omega-\Omega(3',2)) \, ,
\nonumber \\
A(\pi,\omega)&=&(1-\alpha^2)\, \delta (\omega-\Omega(2,1')) + 
\alpha^2\, \delta (\omega-\Omega(3,2)) \, ,
\label{Akw_n100}
\end{eqnarray}
where $\alpha^2=\frac{1}{2}(1+\frac{1}{\sqrt{U^2+(4t_\perp)^2}})$
and 
$\Omega(n,m)$ correspond to the excitation energies 
of the various allowed transitions between a state with $m$
particles to a state with $n$ particles.
Here, $n$, $n'$, $n''$, etc... index the GS and the excited states with $n$
particles on a rung of increasing energy.
The poles of the spectral functions are given, also for increasing energies, by
\begin{eqnarray}
\Omega(2,1')&=& \frac{1}{2}(U-\sqrt{U^2+(4t_\perp)^2})-t_\perp \, ,
\nonumber \\
\Omega(2,1)&=&  \frac{1}{2}(U-\sqrt{U^2+(4t_\perp)^2})+t_\perp    \, ,
\\
\Omega(3,2)&=&  \frac{1}{2}(U+\sqrt{U^2+(4t_\perp)^2})-t_\perp    \, , 
\nonumber \\
\Omega(3',2)&=&  \frac{1}{2}(U+\sqrt{U^2+(4t_\perp)^2})+t_\perp     \, .
\nonumber 
\end{eqnarray}
The chemical potential $\mu$ lies between $\Omega(2,1)$ and $\Omega(3,2)$
leading to the same integrated spectral weight ($=1$) in the photoemission and
inverse photoemission parts of the spectrum for all values of $U$. 
Hence, as expected, the system is an insulator with a single particle 
gap of $\Delta_{eh}=\sqrt{U^2+16t_\perp^2}-2t_\perp$.
However, since the weight $\alpha^2$ varies strongly with the ratio 
$U/4t_\perp$, the distribution of spectral weight changes qualitatively for
increasing $U$ as shown in Figs.~\ref{Akw_rung}(a),(b).
At small $U$, $\alpha^2\sim 1-\frac{1}{4}(\frac{U}{4t_\perp})^2$ and
one recovers, as in the non-interacting limit, two highly weighted 
bonding (at an energy around $-t_\perp$) and antibonding (at an energy 
around $t_\perp$) bands. When $U\rightarrow\infty$, $\alpha^2\rightarrow 1/2$
and, thus, with increasing $U$ spectral weight is transferred to bonding and 
antibonding states further away from the chemical potential.
In the large $U/4t_\perp$ regime, the system becomes a four-band insulator 
with 4 (almost) equally weighted poles and a Hubbard gap of order $U$ 
separating bonding or anti-bonding states at $\omega<\mu$ and $\omega>\mu$.

Note that a similar transition from a two-band to a four-band insulator 
has also been observed in QMC studies of the half-filled Hubbard 
ladder~\cite{Hubbard_neq1} at finite $t_\parallel$
and fixed value $U/t_\parallel=8$ by decreasing the ratio 
$t_\perp/t_\parallel$ from 2 to 0.5. In fact, $U/t_\parallel=8$ and 
$t_\perp/t_\parallel=2$ correspond to the intermediate regime
$U/4t_\perp=1$ where, according to the previous $t_\parallel\rightarrow 0$
estimate, only $\sim 15\%$ of the spectral weight is located in the side bands.
For smaller $t_\perp$, more weight appears at the position of these
two additional structures leading to four bands. 
In general, for arbitrary ratio $t_\perp/t_\parallel$, one expects a
transition to a  four-band insulator when $U$ becomes sufficiently
large compared to
the {\it largest} of the two hopping matrix elements i.e. 
$U\gg {\rm Max}\{t_\perp,t_\parallel \}$.

Let us now turn to the discussion of the 
quarter-filled case $n=0.5$ where a similar local
rung calculation leads to, 
\begin{eqnarray}
A(0,\omega)&=&\frac{1}{2}\, \delta (\omega-\Omega(1,0)) 
+ \frac{1}{2}\alpha^2 \, \delta (\omega-\Omega(2,1))
\nonumber \\
&+&\frac{1}{2}(1-\alpha^2)\, \delta (\omega-\Omega(2^{'''},1)) \, ,
\label{Akw_n050} \\
A(\pi,\omega)&=&\frac{3}{4}\, \delta (\omega-\Omega(2',1)) + 
\frac{1}{4}\, \delta (\omega-\Omega(2'',1)) \, ,
\nonumber 
\end{eqnarray}
where the new energy poles are given by
\begin{eqnarray}
\Omega(1,0)&=& -t_\perp \, ,
\nonumber \\
\Omega(2',1)&=&  t_\perp    \, ,
\\
\Omega(2'',1)&=&  t_\perp + U   \, , 
\nonumber \\
\Omega(2''',1)&=&\Omega(3',2)   \, .
\nonumber 
\end{eqnarray}
Since the chemical potential is located exactly {\it between} 
$\Omega(1,0)$ and $\Omega(2',1)$, the system is an insulator for 
all values of $U$ and (sufficiently) small $t_\parallel$ (compared to $U$). 
However, $A(q_\perp,\omega)$ 
exhibits completely different forms at small and large $U$ couplings
as observed in Figs.~\ref{Akw_rung}(c),(d). 
At small $U$, as in the non-interacting case, the bonding states and
antibonding states are separated by an energy of order $2t_\perp$.
However, each structure is split by a small energy proportional to $U$
and the chemical potential lies between the two $q_\perp=0$ sub-bands.
For large $U$, the gap becomes of order $2t_\perp$ and upper Hubbard
bands (of almost equal weights) of the bonding
and anti-bonding states appear at an energy $\sim U$ higher.
Although this picture does not take into account $t_\parallel$,
the role of a small $t_\parallel$
can be easily discussed qualitatively.
In fact $t_\parallel$ is expected to give a dispersion in the chain direction
and a width to the various structures discussed here. 
When $4t_\parallel$ becomes comparable to the single particle excitation 
gap $\Delta_{eh}$, bands will start to overlap and 
a transition from the insulator to a metallic state is 
expected~\cite{note1}, as will be discussed in 
the next section. Since the single particle excitation 
gap $\Delta_{eh}=\frac{1}{2}(U-\sqrt{U^2+16t_\perp^2})+2t_\perp$ is of the 
order of the smallest of the two energy scales $U/2$ and $2t_\perp$,
the insulating phase is then restricted to the range 
$4t_\parallel < {\rm Min}\{ U/2,2t_\perp\}$. 

The existence of a metal-insulator transition is, in fact, specific to 
quarter filling (besides the half-filled case which is always insulating).
A simple argument is here presented to show that 
at other (commensurate) densities such as $n=0.75$ the metallic phase (i.e.
with at least one gapless charge mode) is 
stable for arbitrary small $t_\parallel$.
At $n=0.75$, a local rung calculation of $A(q_\perp,\omega)$ requires 
to consider as a GS two decoupled rungs on 4 sites with 2 and 1 particle,
respectively. The spectral function is then given straightforwardly 
by the average of the spectral function Eqs.(\ref{Akw_n100}) and (\ref{Akw_n050})
at densities $n=1$ and $n=0.5$. However, the location of the chemical potential
is a subtle issue: since the states at the energy $\omega=\Omega(2,1)$ 
are completely filled (empty) for $n=1$ ($n=0.5$), it is clear that
this state will become {\it partially} filled at $n=0.75$ so that the 
chemical potential is pinned at this energy. Consideration of the 
spectral weights of the excitations shows immediately that, for
arbitrary small $t_\parallel$, the band centered at $\omega=\Omega(2,1)$ 
(of weight $\frac{3}{4}\alpha^2$) is always $\frac{2}{3}$--filled 
leading to a metallic behavior. 
In this case, an additional interaction, e.g. between
nearest neighbor sites along the chains, would be required to
produce a metal-insulator transition.
Schematic representations of $A(q_\perp,\omega)$ at $n=0.75$ are shown
in Figs.~\ref{Akw_rung}(c,f) at small and large $U$. 
At small $U$, as expected, the two bonding and anti-bonding structures 
separated by $\sim 2t_\perp$ are clearly visible and the bonding
states at the lower energies are partially occupied. In this limit,
$U$ leads essentially to small splittings of the various structures
into sub-bands (as for $n=0.5$).
For large $U$, the spectral function is qualitatively very different
with 2 distinct bands around $-t_\perp$ and $t_\perp$ for both $q_\perp=0$
and $q_\perp=\pi$ states. However, the upper Hubbard band around an
energy $\sim U$ is formed of two peaks (separated by $2t_\perp$) for
$q_\perp=\pi$ while only one peak is present for $q_\perp=0$.

Finally in this section a brief discussion of the case of the
$t-J$ ladder is included. Since this model describes only the low energy
properties
of the Hubbard model,
the corresponding spectral functions in the $t_\parallel\rightarrow 0$
limit can be obtained easily from the previous ones by discarding the
high energy peaks whose energy scales as $U$ for large--$U$, 
setting $\alpha^2=1/2$ and expanding energies to first order in 
$J_\perp=4t_\perp^2/U$.
In fact, it can be easily shown that the same expressions hold for 
the $t-J$ model (with arbitrary $J_\perp$).
Note that, due to the projection of the high energy states, the 
spectral function of the $t-J$ model follows the new sum-rule
$\int A(q_\perp,\omega)\, d\omega=\frac{1+x}{2}$ (instead of 1), 
where $x=1-n$ is the doping fraction.
At half-filling one gets:
\begin{eqnarray}
A(0,\omega)&=&\frac{1}{2}\, \delta (\omega-(t_\perp-J_\perp))  \, ,
\nonumber \\
A(\pi,\omega)&=&\frac{1}{2}\, \delta (\omega-(-t_\perp-J_\perp)) \, ,
\label{AkwtJ_n100}
\end{eqnarray}
with the chemical potential located at an higher energy ($\sim U/2$).
Similarly, at quarter-filling one obtains,
\begin{eqnarray}
A(0,\omega)&=&\frac{1}{2}\, \delta (\omega-(-t_\perp)) +
\frac{1}{4}\, \delta (\omega-(t_\perp - J_\perp)) \, ,
\nonumber \\
A(\pi,\omega)&=&\frac{3}{4}\, \delta (\omega - t_\perp) \, ,
\label{AkwtJ_n050}
\end{eqnarray} 
with the chemical potential located between $-t_\perp$ and 
$t_\perp - J_\perp$. 
It is interesting to notice that, when $J_\perp$ exceeds
$2t_\perp$, the electron-like excitation becomes lower in energy than
the hole-like excitation. This signals the onset of phase
separation or, alternatively, some sort of charge localization/ordering
(such as charge density wave ordering). Physically, this
occurs when the magnetic energy gain of a singlet on a single rung becomes
larger than the kinetic energy of two particles on individual rungs.

\subsection{Exact diagonalization results: Hubbard model}

Let us now investigate the dynamical properties of 
the Hubbard and $t-J$ models for arbitrary parameters 
using exact diagonalization techniques. 
Cyclic $2\times L$ ladders are diagonalized and
the (zero temperature)
particle spectral function is obtained exactly by a standard 
continued-fraction procedure.
Although in practice one is limited to $L=8$ (for the Hubbard model), 
both periodic (PBC) and anti-periodic (ABC) boundary
conditions can be used to consider
a sufficiently large number of momenta $q_\parallel=n\frac{\pi}{L}$,
$n=0,2L-1$.
 
\begin{figure}[htbp]
\begin{center}
\vspace{-0.65truecm}
\psfig{figure=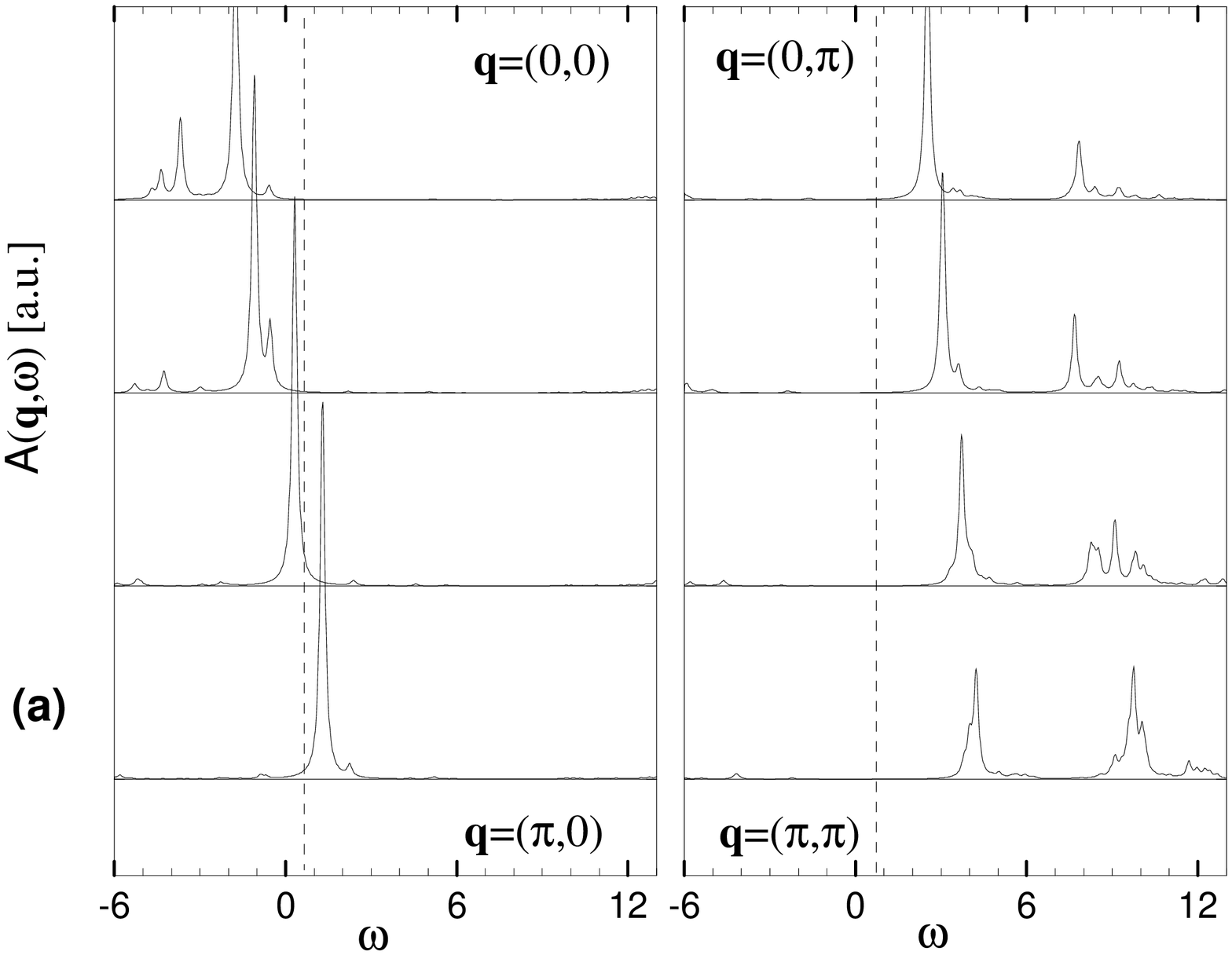,width=8.6truecm,angle=-90}
\vspace{-0.65truecm}
\psfig{figure=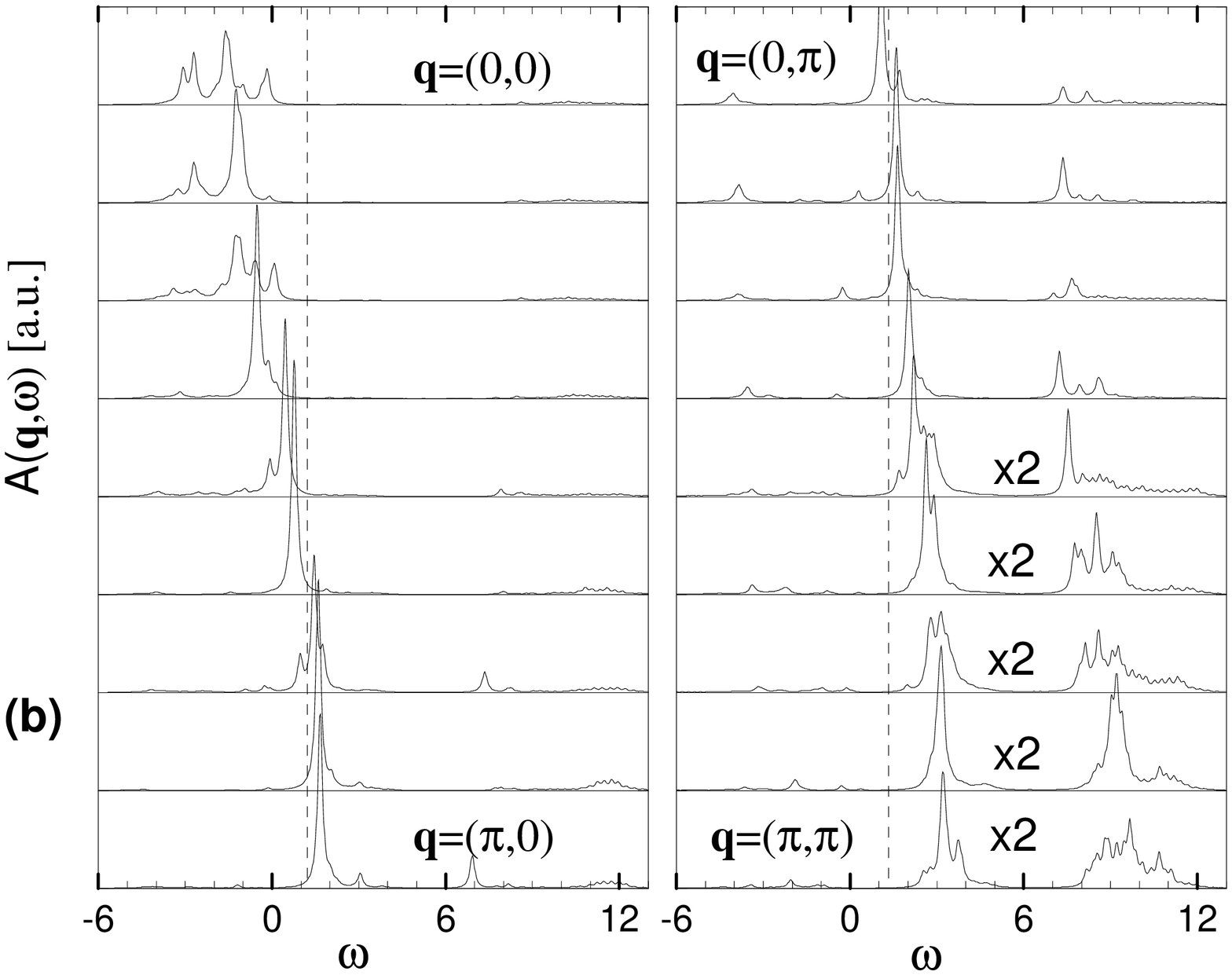,width=8.6truecm,angle=-90}
\vspace{-0.65truecm}
\psfig{figure=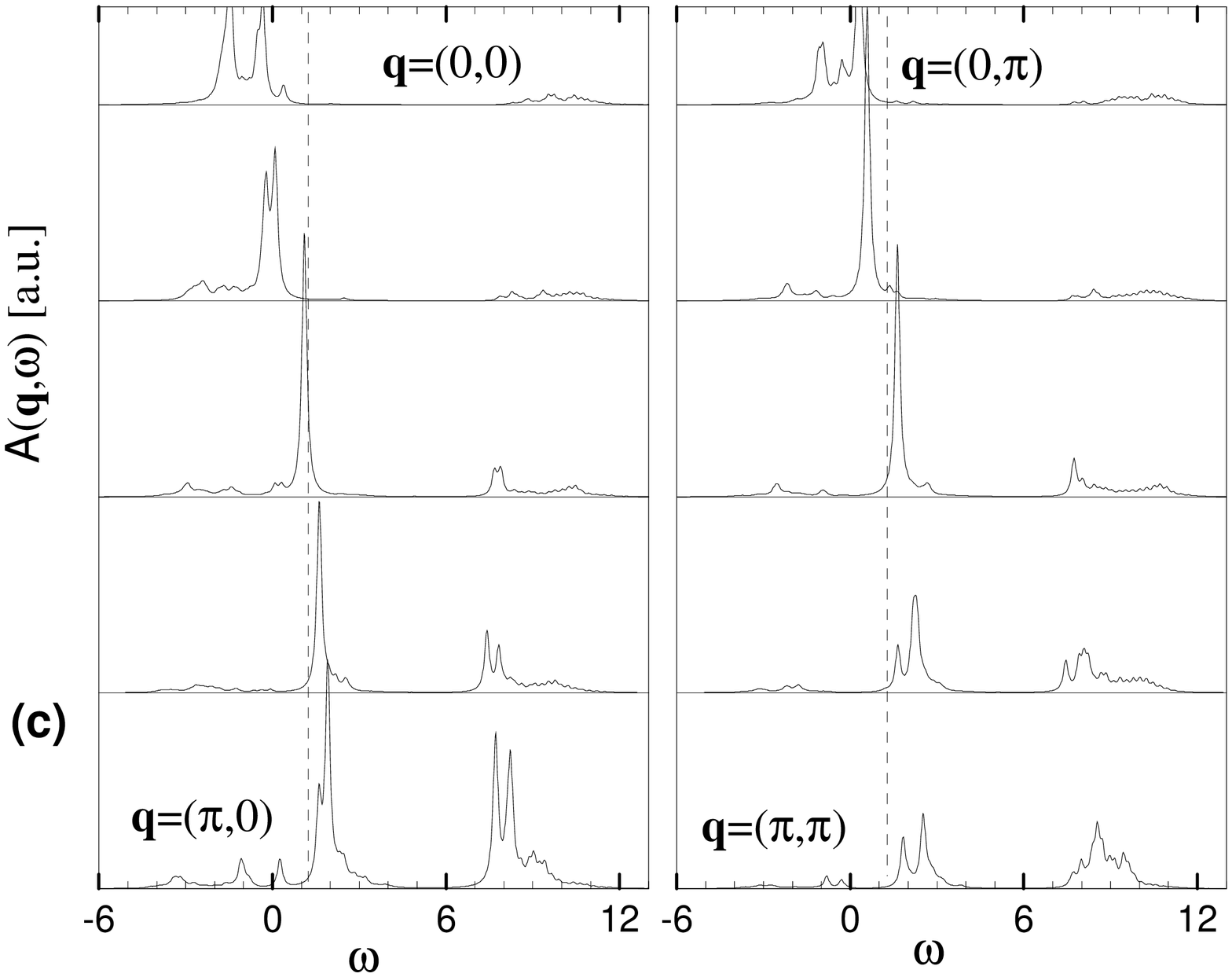,width=8.6truecm,angle=-90}
\end{center}
\caption{
Spectral function $A({\bf q},\omega)$ of the Hubbard ladder for $U=8$ and
$n=0.75$. The left 
and right sides correspond to the bonding ($k_y=0$) and anti-bonding
states ($k_y=\pi$), respectively, and $k_x$ runs from 0 to $\pi$ from 
the top to the bottom. The position of the chemical
potential is indicated by a vertical dotted line.
(a), (b) and (c) correspond to $t_\perp=2.5$,
$t_\perp=1.5$ and $t_\perp=0.5$, respectively. 
}
\label{Akw1_Hubbard}
\end{figure}

The case of the Hubbard ladder will be considered first, before focusing
on the low energy excitations described by the $t-J$ model.
The spectral function $A({\bf q},\omega)$ at a density of
$n=0.75$ is shown in Figs.~\ref{Akw1_Hubbard}(a,b,c) for 
$U=8$ and several values of $t_\perp$ ranging from 
2.5 down to 0.5. 
Note that both PBC and ABC have been used in Fig.~\ref{Akw1_Hubbard}(b) while,
in order to reduce CPU time, only ABC (PBC) have been used 
in Fig.~\ref{Akw1_Hubbard}(a) (Fig.~\ref{Akw1_Hubbard}(c)).
Two sharp structures separated by an energy 
proportional to $t_\perp$ 
can be attributed to a bonding and an anti-bonding band.
At the largest ratio of $t_\perp/t_\parallel=2.5$ that  have been considered,
the spectrum exhibits some features of Fig.~\ref{Akw_rung}(c)
obtained in the local rung approximation at small coupling:
(i) in the photoemission part, a $q_\perp=0$ sub-band of small 
spectral weight can be observed at an
energy of about $U/2$ from the main $q_\perp=0$ band crossing 
the chemical potential; (ii) a $q_\perp=\pi$ upper Hubbard band appears 
at an energy $\propto U$ away from the main (empty) $q_\perp=\pi$ band.
On the other hand, some tiny structures characteristic of the 
strong coupling limit (Fig.~\ref{Akw_rung}(f)) can also be 
observed: (i) a small spectral weight exists at $\omega<\mu$ (around 
$\omega\sim -5$) for $q_\perp=\pi$ together with (ii) a quite small 
$q_\perp=0$ upper Hubbard band at $\omega >\mu$. Interestingly enough, these
features become more important for $t_\perp=1.5$ as
shown in Fig.~\ref{Akw1_Hubbard}(b) which corresponds, in fact, to a 
larger ratio $U/4t_\perp \simeq 1.3$. 

With decreasing electron density, the respective 
position of the two main bands 
and the position of the Fermi level seems to evolve as in a rigid-band 
scheme. However, there are important differences:
(i) the bandwidth is strongly reduced specially at smaller 
$t_\perp/t_\parallel$;
(ii) the excitations become sharper when the band
crosses the chemical potential. 
To the best of our knowledge, this is the first observation in a numerical 
study of the broadening of the ``quasi-particle''--like peaks 
excitations as one moves away from the chemical potential.

\begin{figure}[htbp]
\begin{center}
\vspace{-0.65truecm}
\psfig{figure=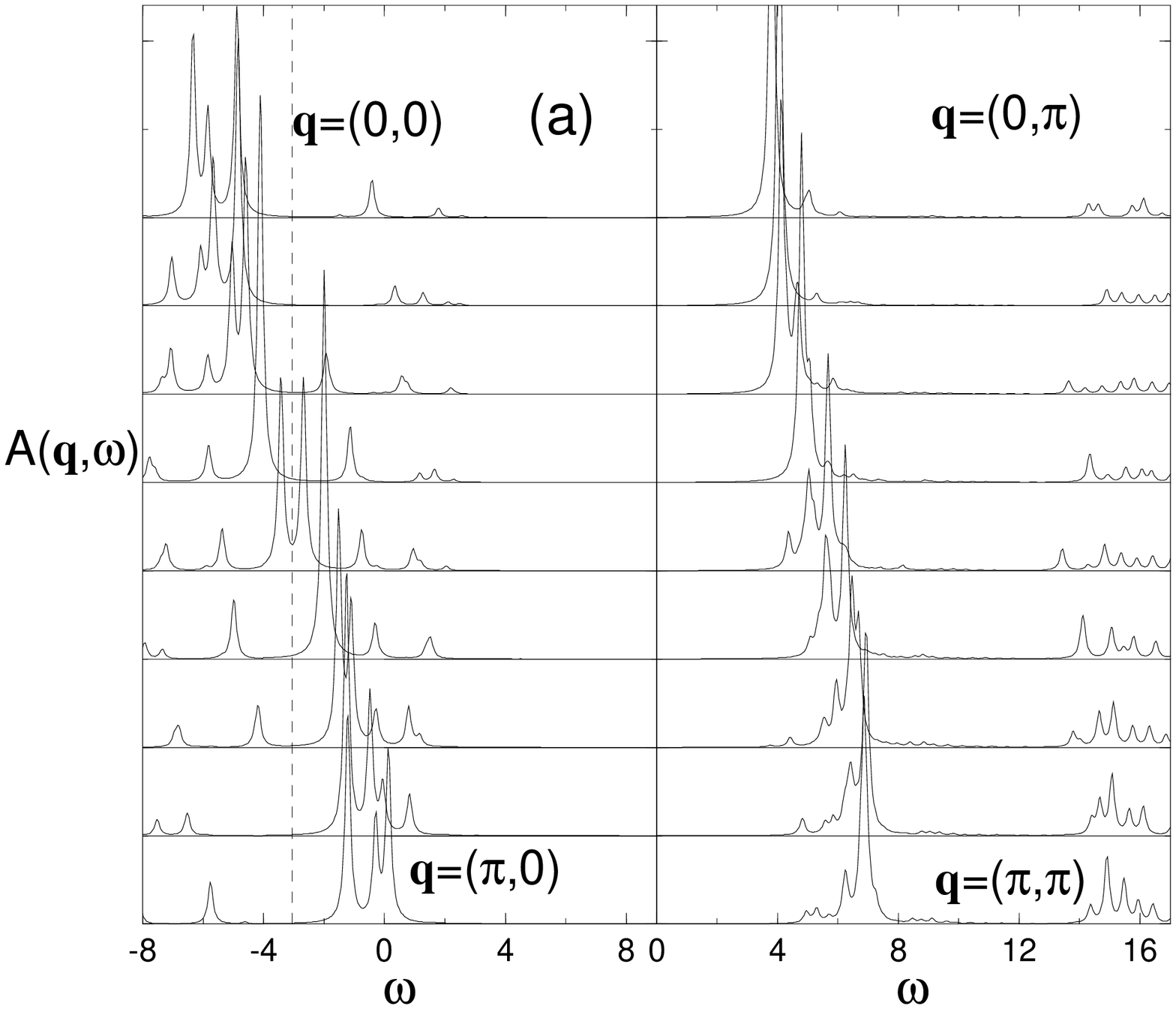,width=8.6truecm,angle=-90}
\vspace{-0.65truecm}
\psfig{figure=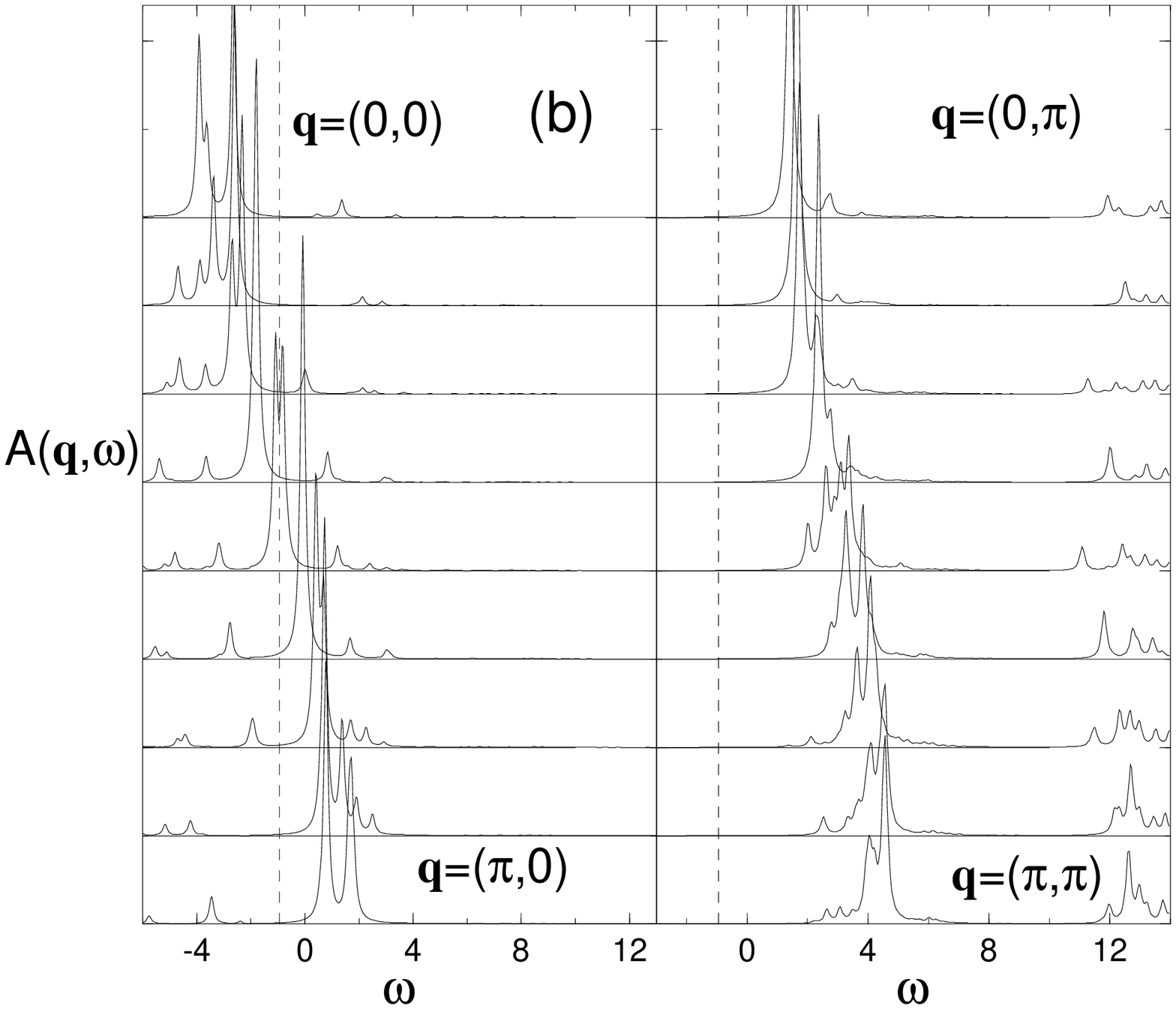,width=8.6truecm,angle=-90}
\vspace{-0.65truecm}
\psfig{figure=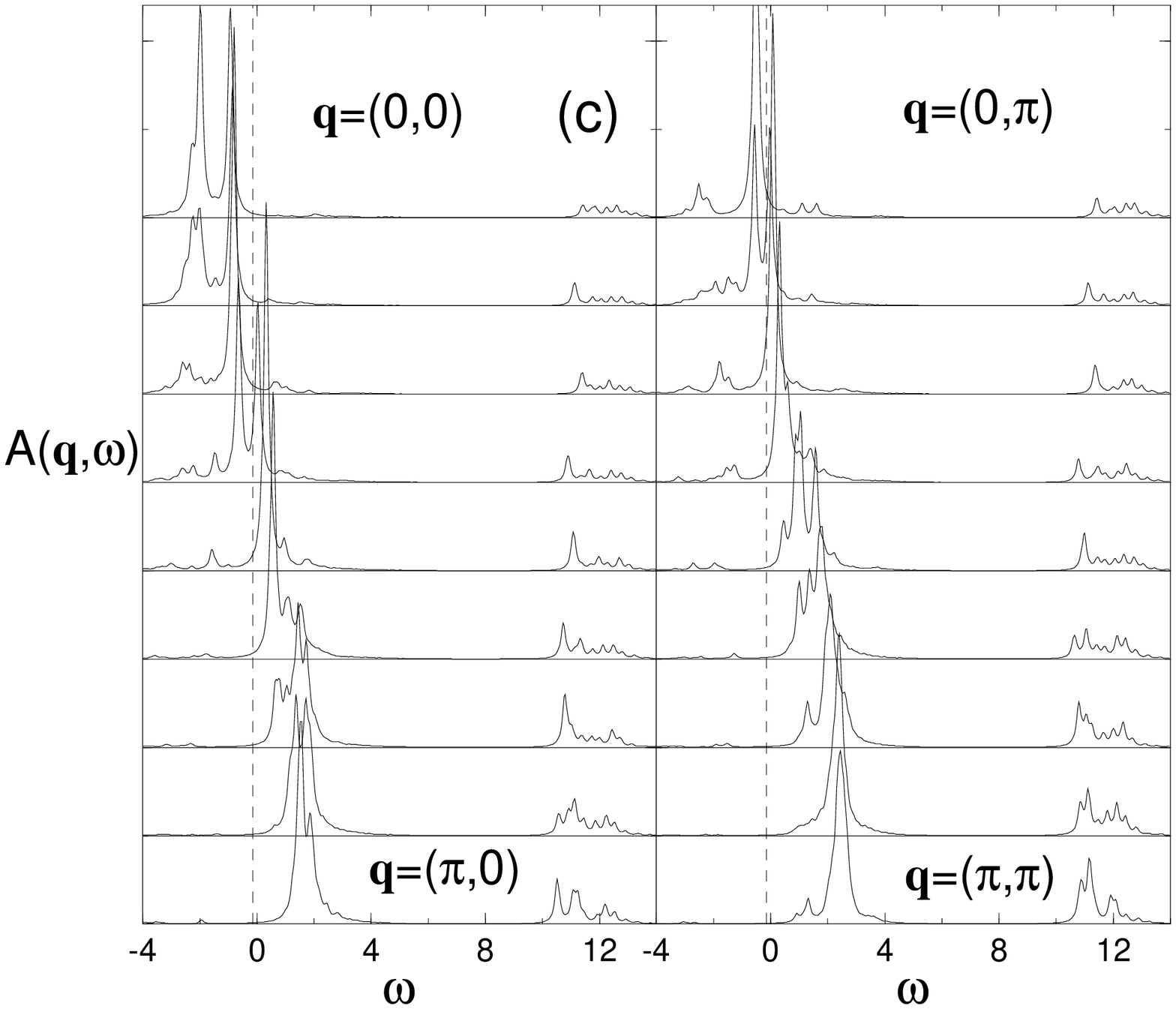,width=8.6truecm,angle=-90}
\end{center}
\caption{
Spectral function $A({\bf q},\omega)$ of the Hubbard ladder for $U=10$
at quarter-filling $n=0.5$. The left 
and right sides correspond to the bonding ($k_y=0$) and anti-bonding
states ($k_y=\pi$), respectively, and $k_x$ runs from 0 to $\pi$ from 
the top to the bottom. (a), (b) and (c) correspond to $U=10$ with
$t_\perp=5$, $t_\perp=2.5$ and $t_\perp=0.5$ respectively. 
}
\label{Akw2_Hubbard}
\end{figure}

For a larger hole doping and working  at a commensurate value of $n=0.5$
qualitative changes can take place in the spectral function
at sufficiently large $U$ and $t_\perp$.
Data are shown in Fig.~\ref{Akw2_Hubbard}. For $t_\perp=0.5$ the
two partially filled bonding and antibonding bands can be observed together
with their corresponding upper Hubbard bands at higher energy.
As expected from the previous 
$t_\parallel\rightarrow 0$ analysis, the spectral weight of the 
$q_\perp=0$ upper Hubbard band, at fixed $U$, gets strongly reduced for 
increasing $t_\perp$ i.e. for a decreasing ratio
$U/t_\perp$.
At large enough $t_\perp$,  
a gap appears in the $q_\perp=0$ structure, leading
to two sub-bands and an insulating behavior in agreement with the
local rung calculation. 
Such a metal-insulator transition is induced by a
combined effect of $t_\perp$ and $U$: when $t_\perp$ is large enough 
the lower band becomes half-filled and a finite U, leading to
relevant Umklapp scattering, can then open a gap. 

\subsection{Exact diagonalization results: $t-J$ model}

\begin{figure}[htbp]
\begin{center}
\vspace{-0.3truecm}
\psfig{figure=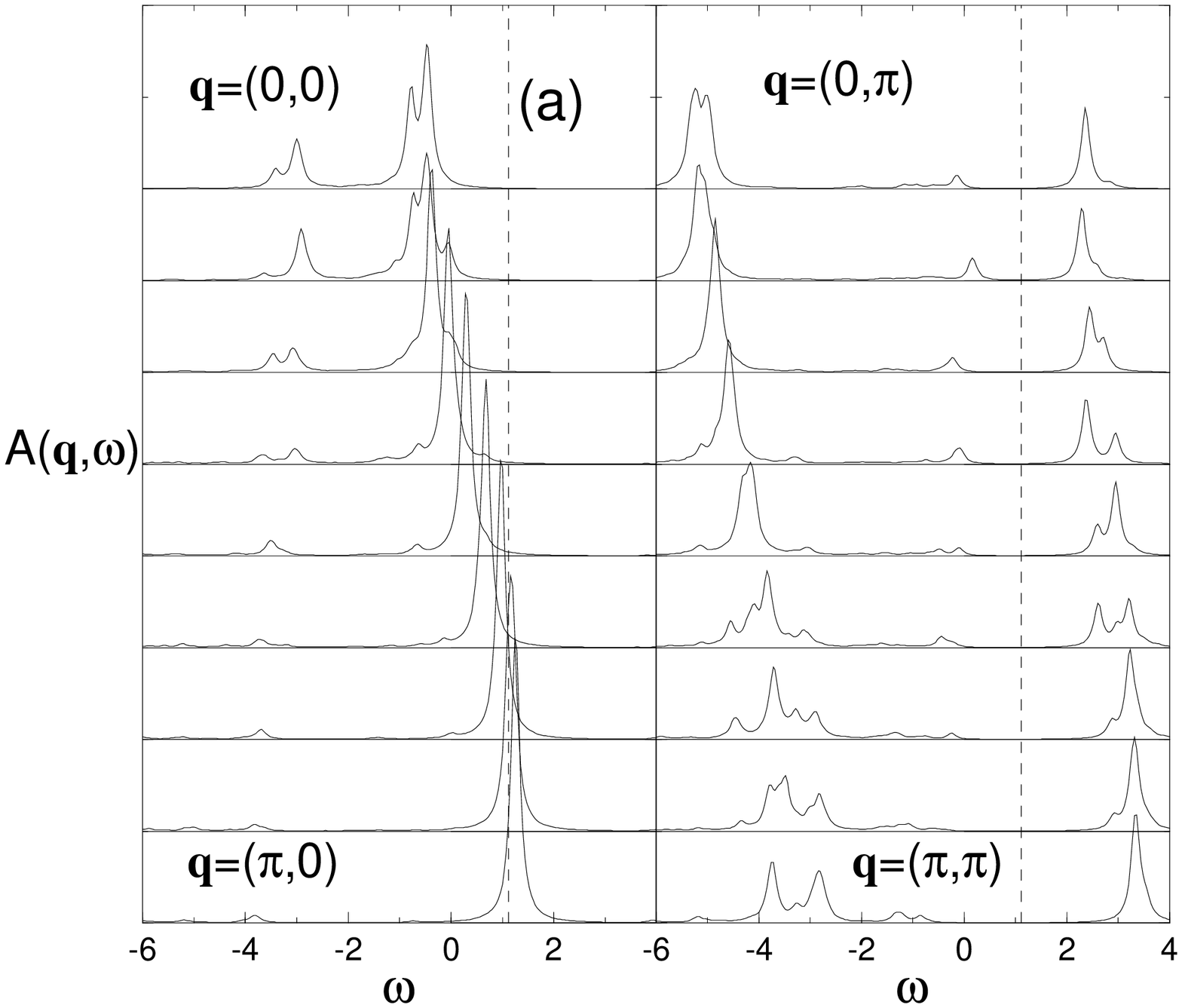,width=8.6truecm,angle=-90}
\vspace{-0.3truecm}
\psfig{figure=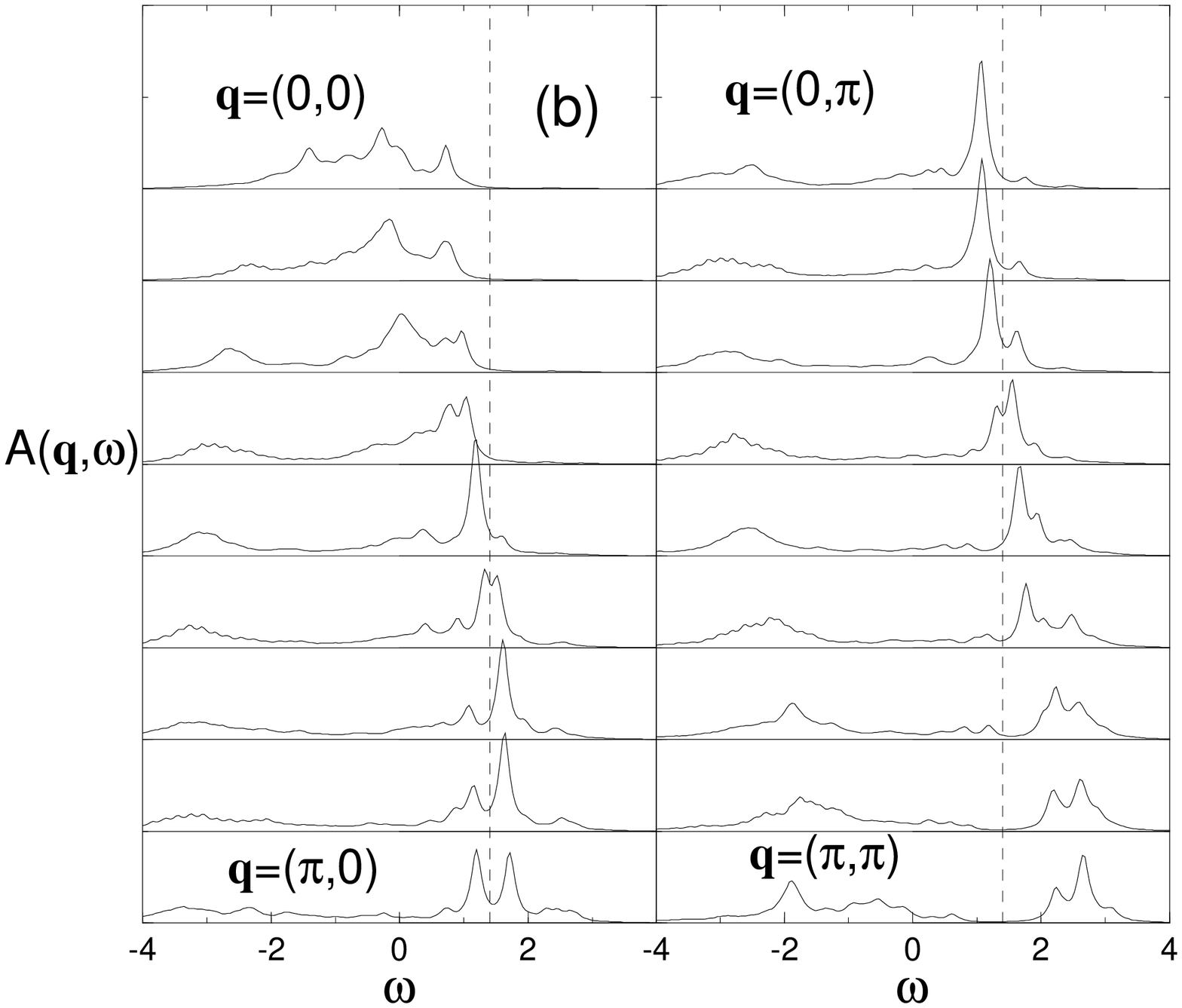,width=8.6truecm,angle=-90}
\vspace{-0.3truecm}
\psfig{figure=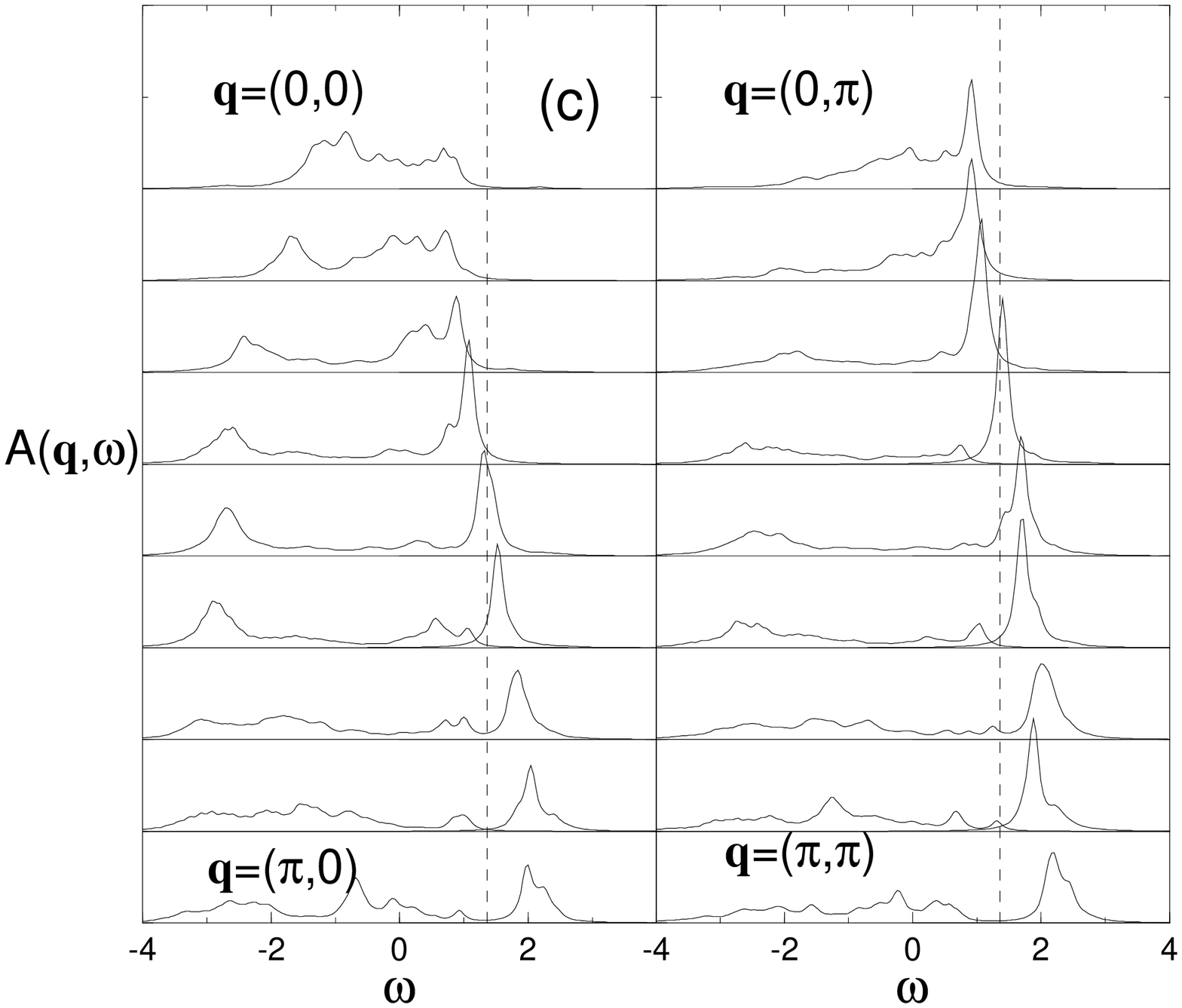,width=8.6truecm,angle=-90}
\end{center}
\caption{
Spectral function $A({\bf q},\omega)$ of the $t-J$ ladder at $n=0.875$
and $J_\parallel=0.4$. 
Conventions are similar to those of Fig. \protect\ref{Akw1_Hubbard}.
(a), (b) and (c) correspond to $t_\perp=2$,
$t_\perp=1$ and $t_\perp=0.5$, respectively. 
}
\label{Akw1_tJ}
\end{figure}

\begin{figure}[htbp]
\begin{center}
\vspace{-0.3truecm}
\psfig{figure=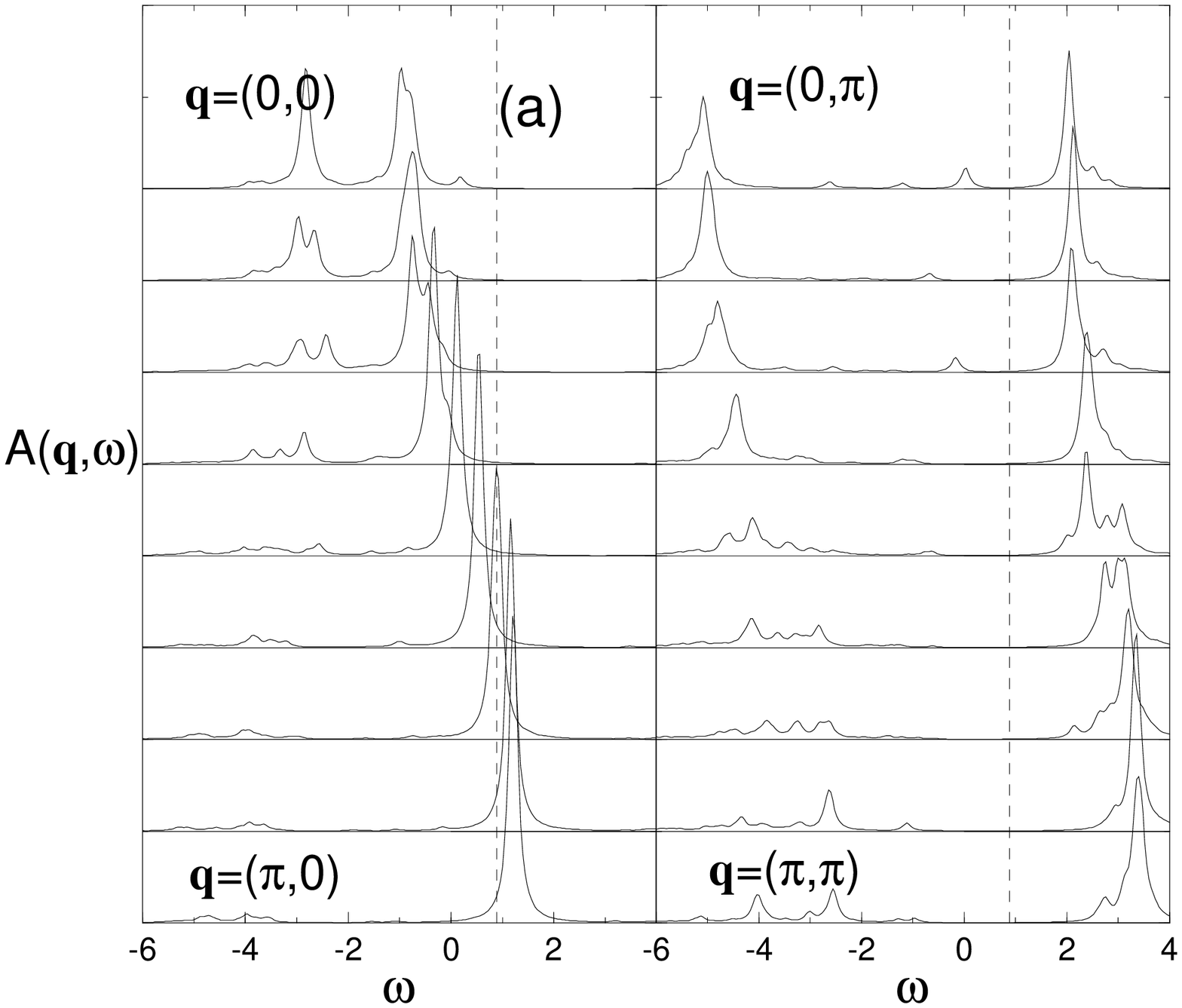,width=8.6truecm,angle=-90}
\vspace{-0.3truecm}
\psfig{figure=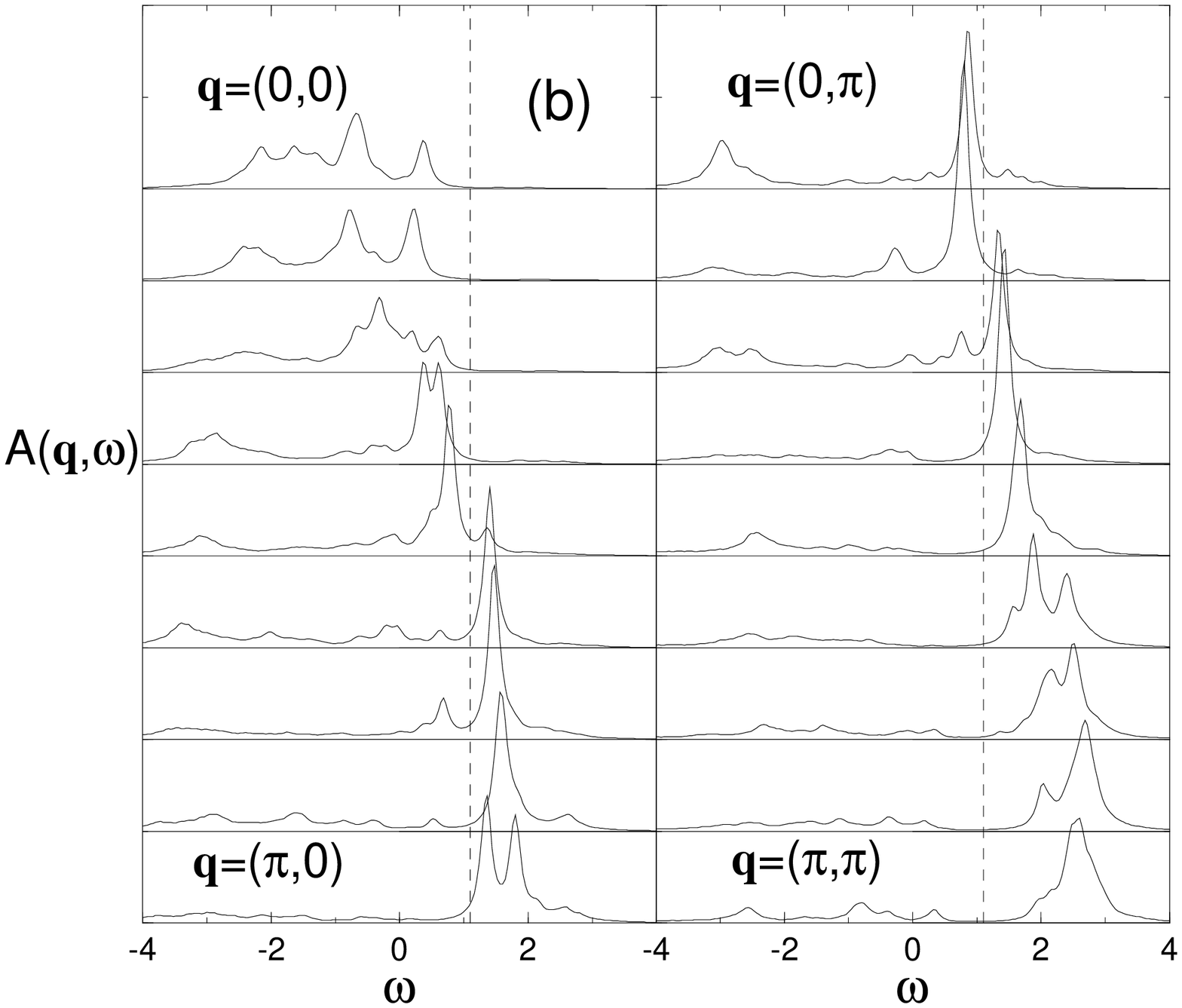,width=8.6truecm,angle=-90}
\vspace{-0.3truecm}
\psfig{figure=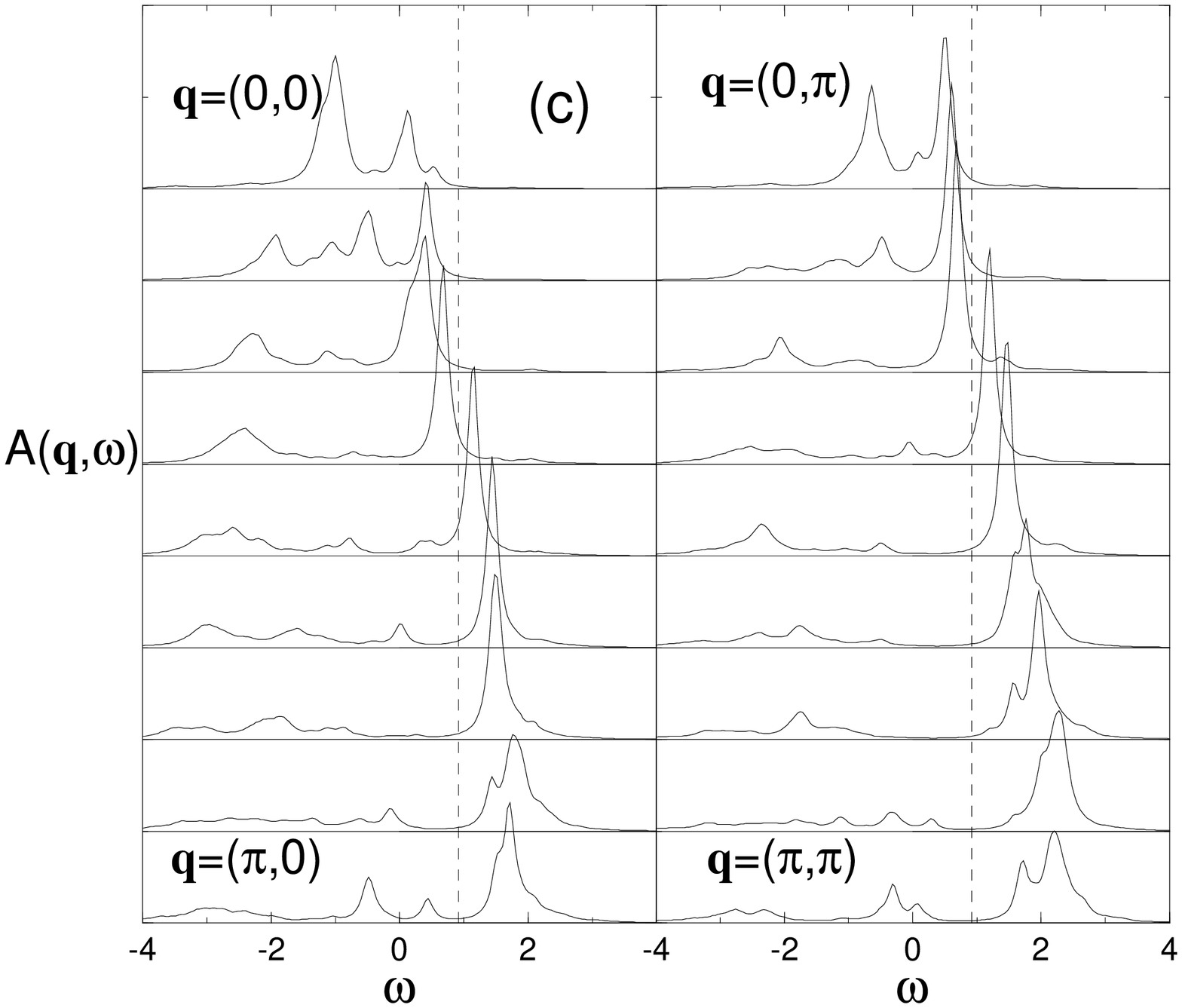,width=8.6truecm,angle=-90}
\end{center}
\caption{
Spectral function $A({\bf q},\omega)$ of the $t-J$ ladder at $n=0.75$
and $J_\parallel=0.4$. 
Conventions are similar to those of Fig. \protect\ref{Akw1_Hubbard}.
(a), (b) and (c) correspond to $t_\perp=2$,
$t_\perp=1$ and $t_\perp=0.5$, respectively. 
}
\label{Akw2_tJ}
\end{figure}

In order to study more precisely the influence of doping 
at small energy scales let us now
focus on the $t-J$ ladder~\cite{Haas}. 
Results at small hole densities
$n=0.875$ and $n=0.75$ are shown in Figs.~\ref{Akw1_tJ}
and Figs.~\ref{Akw2_tJ} and are 
consistent with the previous results on the Hubbard model.

\begin{figure}[htbp]
\begin{center}
\vspace{-0.3truecm}
\psfig{figure=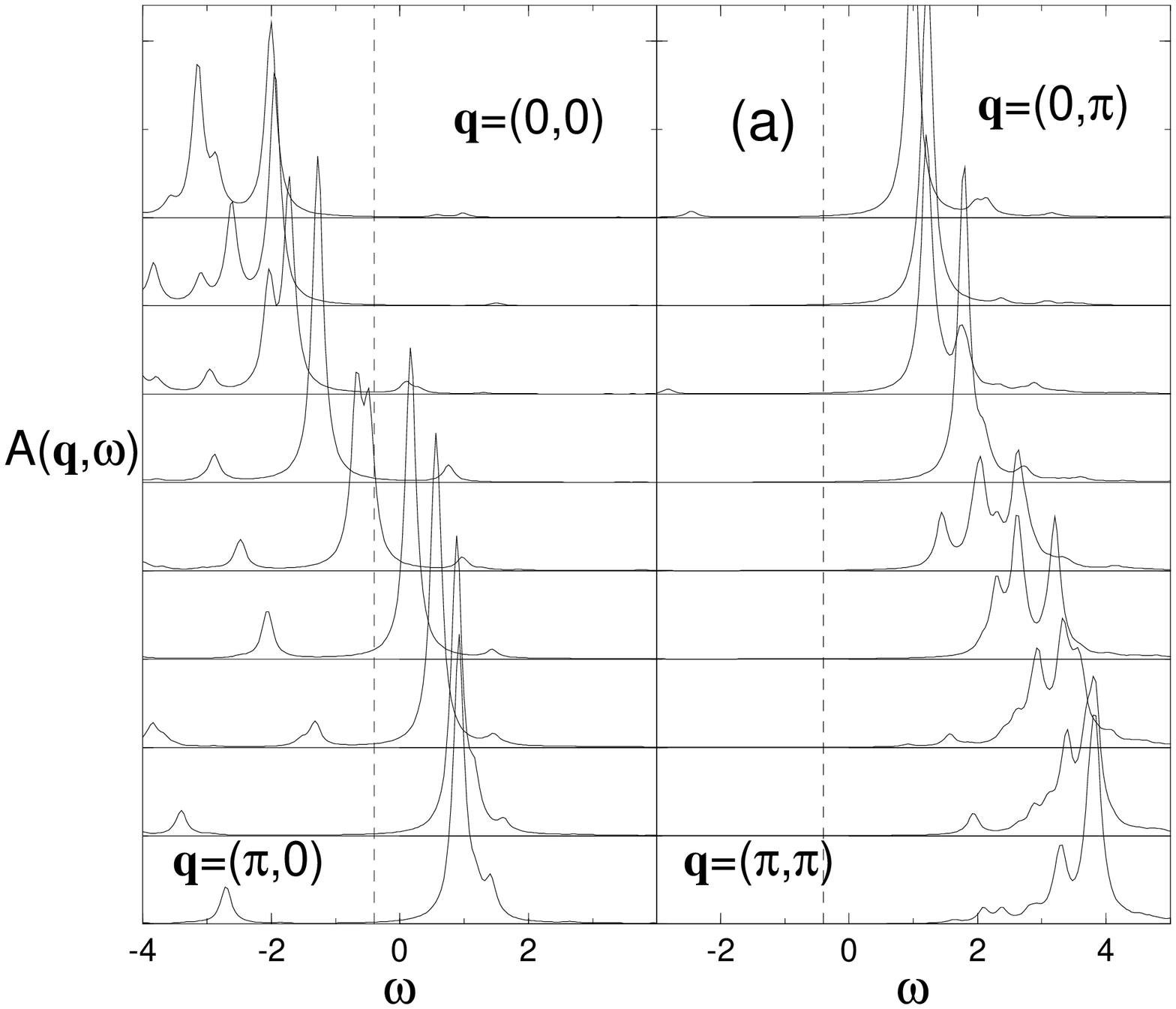,width=8.6truecm,angle=-90}
\vspace{-0.3truecm}
\psfig{figure=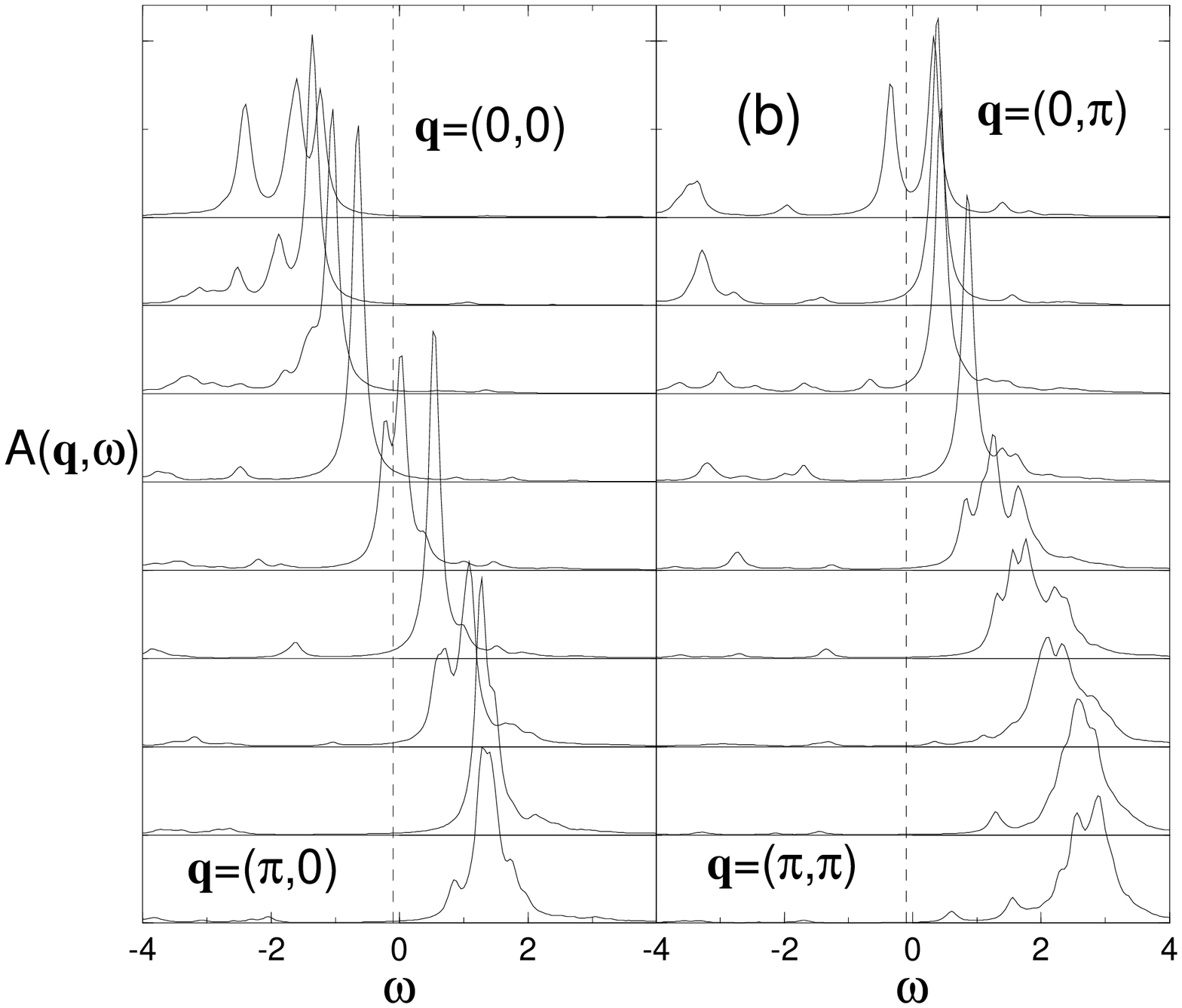,width=8.6truecm,angle=-90}
\vspace{-0.3truecm}
\psfig{figure=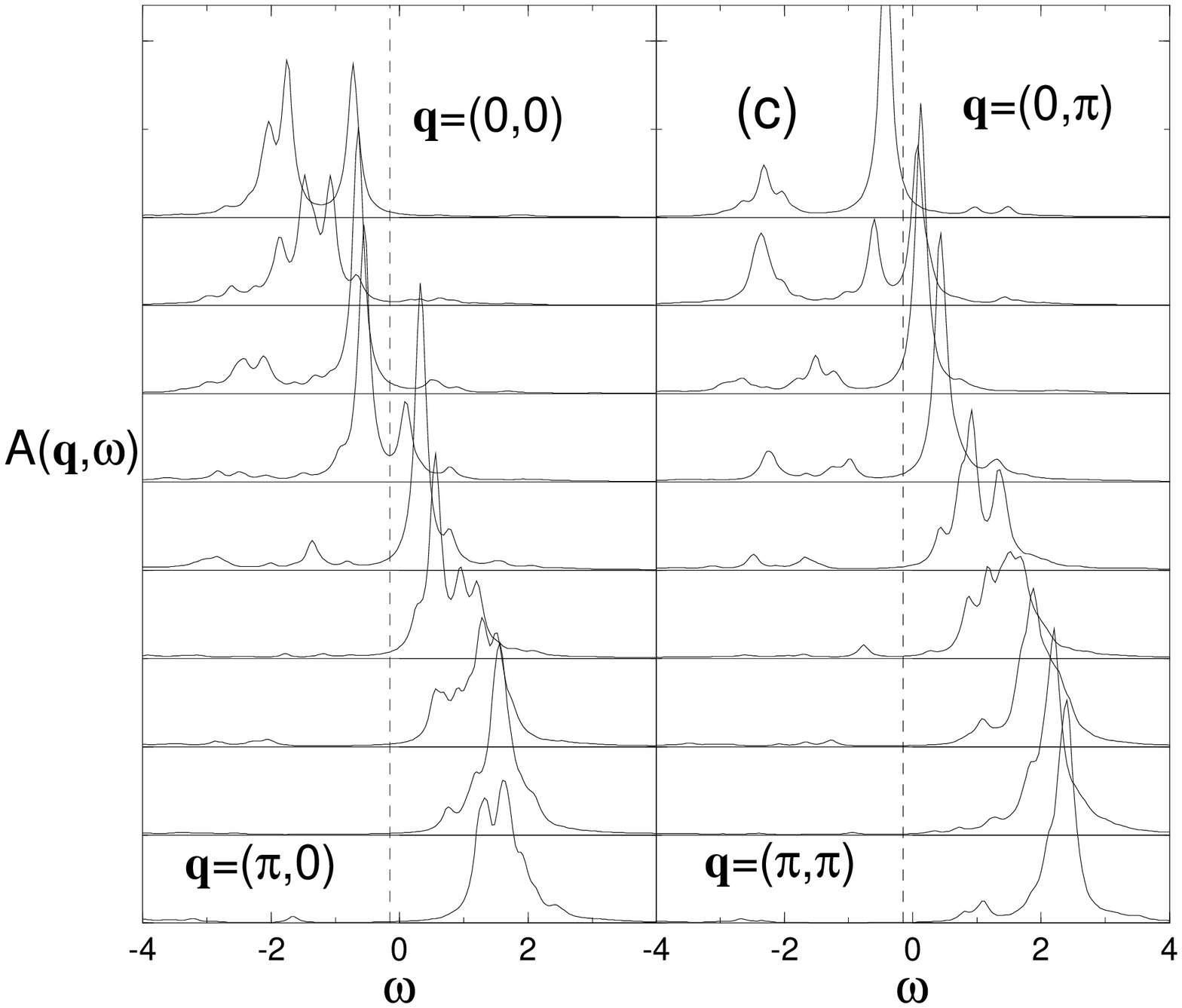,width=8.6truecm,angle=-90}
\end{center}
\caption{
Spectral function $A({\bf q},\omega)$ of the $t-J$ ladder at $n=0.5$
and $J_\parallel=0.2$. 
Conventions are similar to those of Fig. \protect\ref{Akw1_Hubbard}.
(a), (b) and (c) correspond to $t_\perp=2$,
$t_\perp=1$ and $t_\perp=0.5$, respectively. 
}
\label{Akw3_tJ}
\end{figure}

Let us first discuss the role of the hole doping $x=1-n$, for the largest
value of $t_\perp=2$  considered here (see Fig.~\ref{Akw1_tJ}(a) and
Fig.~\ref{Akw2_tJ}(a)). For this choice of  parameters, the
$x$-dependence can be qualitatively understood from the single rung picture.
For $t_\parallel\rightarrow 0$ the GS contains a density of $2x$ 
singly occupied bonds and $1-2x$ doubly occupied bonds. By combining the
spectral functions at $n=0.5$ and $n=1$ with the respective weights,
one obtains a simple picture of the influence of doping consistent with
the numerical results at small (but finite) $t_\parallel$.
The $q_\perp=0$ main structure (which is the closest to the chemical potential
at half-filling) becomes partially filled with a weight of $x/2$ in the 
inverse photoemission part $\omega>\mu$. Note that the dispersion of the band
is especially flat in the vicinity of the chemical potential at small $x$.
With increasing doping, weight 
is transferred from this structure (of total weight $1/2-x/2$) and from the 
upper Hubbard band (not described by the $t-J$ model) to $q_\perp=0$
states further away from the chemical potential. This leads to an emerging
structure of weight $x$ at an energy of $\sim 2t_\perp-J_\perp$ below the
main band. Physically, in a photoemission experiment,
these small peaks correspond to processes where an electron on a 
singly occupied rung is removed by a photon and leaves behind an empty rung.
Note that this structure becomes particularly strong at quarter filling
(as seen in  Fig.~\ref{Akw3_tJ}(a)) where it carries $1/2$ 
of the total spectral weight (normalized to $1$).
In the $q_\perp=\pi$ sector, the main structure in the photoemission
part of the spectrum $\omega<\mu$ (barely seen in the case of the Hubbard model
for the parameters chosen in the previous study) is also loosing 
spectral weight 
upon doping with a total weight of $1/2-x$. The missing weight 
(and some additional spectral weight from the upper Hubbard band) 
is transferred into the inverse photoemission spectrum leading to an emerging
band of total weight $3x/2$ at $\omega>\mu$. 
Such states, obtained by suddenly adding an electron on
a singly occupied rung could be seen in an inverse photoemission experiment. 
At quarter filling $n=0.5$, as seen in Fig.~\ref{Akw3_tJ}(a), the transfer of
spectral weight is complete and the $\omega<\mu$, $q_\perp=\pi$
structure has totally disappeared.

At smaller values of $t_\perp$ (see Figs.~\ref{Akw1_tJ}(b,c) and
Figs.~\ref{Akw2_tJ}(b,c)) the two separate structures, both for $q_\perp=0$
or $q_\perp=\pi$, merge into a single broad structure. 
The data can be fairly well
described by (i) $q_\perp=0$ and $q_\perp=\pi$ bands dispersing 
through the chemical potential and (ii) a broad incoherent background
extending further away from the chemical potential towards negative energies.
Note that, similarly to the previous case of the Hubbard model, the 
peaks of the band-like feature seem to become narrower when they cross the
chemical potential as expected in a Fermi liquid description. 

This Section ends with a short discussion on the possible existence
of small single particle gaps in the previous data. 
The quarter-filled case is qualitatively different from the low doping 
regime. At $n=0.5$ the $t_\parallel\rightarrow 0$ analysis unambiguously
predicts the existence of a gap in the single particle spectrum 
at sufficiently large $t_\perp$ and $U$. 
However, when the $q_\perp=\pi$ structure is not completely empty
(i.e. totally located in the $\omega>\mu$ region of 
the spectrum), as it is the case
in  Figs.~\ref{Akw2_Hubbard}(c)~and~\ref{Akw3_tJ}(b,c), no gap is expected as it is 
clear in the numerical data.
Then, a metal-insulator transition is expected by increasing $t_\perp$ but
whether this transition is driven by $t_\perp$ alone is still unclear.
The data of Figs.~\ref{Akw2_Hubbard}(a,b) corresponding to a
situation where the antibonding band is clearly unoccupied
do not allow to accurately determine a critical value of
$t_\perp$ at which the gap starts to grow.
However, we have checked numerically (not shown) that, by
reducing $t_\parallel$, the spectrum of Fig.~\ref{Akw2_Hubbard}(a) smoothly
evolves into the spectrum obtained above in the single rung approximation,
e.g. exhibiting a well defined gap at the chemical potential. 

At small doping, on the other hand, the physical origin of a
small single particle gap would be quite different. 
In this case, it would be related to the formation of pairs. 
In the $t_\parallel\rightarrow 0$ limit, pairs become stable only when
$J_\perp > 2t_\perp$, i.e. when the magnetic energy on a rung exceeds 
the kinetic energy loss. Otherwise, for $J_\perp < 2t_\perp$, 
the spin gap is immediately destroyed by doping (strictly for 
$t_\parallel=0$) since 
the presence of singly occupied rungs leads to new low-energy spin-1
excitations in the $n=1$ spin gap (of order $J_\perp$).
Therefore, intermediate ratios 
of $t_\perp/t_\parallel$ seem to be more favorable for pair binding.
Although spectra like those shown in Figs.~\ref{Akw1_tJ}(b) and 
Figs.~\ref{Akw2_tJ}(b) are not inconsistent with the presence of
a small gap at the chemical potential, the study of pair binding from 
an investigation 
of the spectral function at small energy scales around the chemical
potential is a difficult task. In order to clarify this issue,
a complementary study of static physical quantities 
is shown in the next Section. 

\section{Superconducting properties}

\subsection{Pair binding energy}

In the  limit where $J_\perp$ is the largest energy scale, 
formation of hole pairs are favored on the rungs
in order to minimize the magnetic energy cost. In fact, this simple naive 
argument breaks down when $t_\perp >J_\perp/2$ since holes on separate
rungs can then benefit from a delocalization on each rung. 
In the large $t_\perp$ limit, a simple 4-sites (2 rungs) calculation
shows that for  $J_\perp/2<<t_\perp$, the pair binding energy (which, as
defined below, should
be positive if a bound state exists) behaves as $J_\parallel^2/
4t_\perp-2t_\parallel$. 
Very tightly bound hole pairs are then not stable in the intermediate regime.
However, extended pairs have been shown to exist in some 
regime~\cite{Dagotto92,pairing,White}.
In this Section,  the stability of the hole pairs as 
a function of the anisotropy ratio will be investigated in detail.

\begin{figure}[htbp]
\begin{center}
\vspace{-0.8truecm}
\psfig{figure=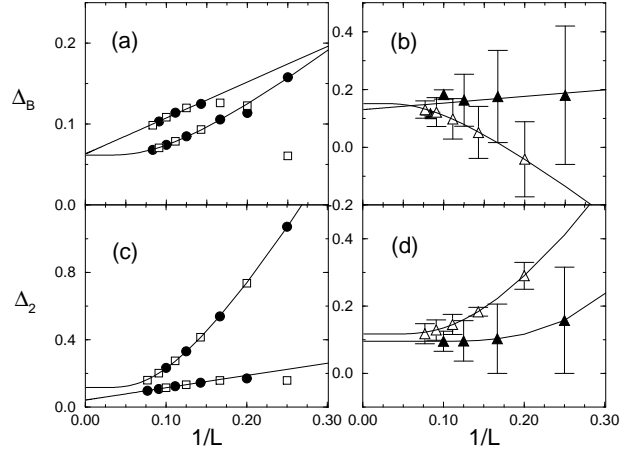,width=8.6truecm,angle=-90}
\end{center}
\caption{
Finite size scaling behaviors as a function of the inverse of the
ladder length for $J_\parallel=0.5$. Filled circles (open squares) correspond 
to PBC (ABC).
The values of $t_\perp$ are shown on the plot.
The full lines correspond to the finite size scaling laws 
used for the extrapolations to $L=\infty$.
(a) Two hole binding energy for $t_\perp =2.25$; (b) Two 
hole binding energy for $t_\perp=1/\sqrt{2}$; 
(c) Finite size behavior of the triplet gap in the GS with 2 holes
for $t_\perp =2.25$;
(d) Finite size behavior of the triplet gap in the GS with 2 holes
for $t_\perp =1/\sqrt{2}$;
In (b) and (d), the triangles correspond to averages between the PBC and the
ABC data and the sizes of the error bars correspond to the
absolute value of the difference between the two sets. Open (filled) symbols
correspond here to $L$ odd (even). 
}
\label{Delta_vs_L}
\end{figure}

To study the onset of pair binding, irrespective of the actual size of
the pair, the two-hole binding energy defined by
\begin{equation}
\Delta_B(L)=2E(L,1)-E(L,2)-E(L,0) \, ,
\end{equation}
is considered,
with  $E(L,N_h)$ being the GS energy of the $2\times L$ ladder doped with
$N_h$ holes.  $E(L,N_h)$ ($N_h=0,1,2$) have been calculated on $t-J$ ladders
with periodic or anti-periodic boundary conditions along the legs direction 
and considering
sizes up to $L=13$. As previously, it has been assumed for convenience that
$J_\perp=J_\parallel(t_\perp/t_\parallel)^2$.
Typical finite size behaviors of $\Delta_B(L)$ are shown 
in Figs.~\ref{Delta_vs_L}(a,b) for different anisotropy ratios.
Some caution is obviously needed in order to extrapolate the results to
the thermodynamic limit. At large $t_\perp$
for a fixed choice of the boundary conditions
very regular oscillations in $\Delta_B$ vs $L$ appear (as observed 
for instance in Fig.~\ref{Delta_vs_L}(a)).
However, the
data set corresponding to ladders with an even (odd) number of rungs ($L$)
and PBC can be combined with the data set corresponding to 
ladders with an odd (even) number of rungs and ABC. 
The two resulting curves exhibit a smooth behavior (full lines) 
which allow 
an accurate extrapolation to the thermodynamic limit. For intermediate
values of $t_\perp/t\sim 1.25$, the two data sets merge into a single curve
and finite size corrections become particularly small. 

For smaller ratios such as $t_\perp/t_\parallel\leq 1$ 
(see Fig.~\ref{Delta_vs_L}(b)),
there is a qualitative change of behavior. In this case, the data
corresponding to ladders with an {\it odd} number of rungs ($L=2p+1$) 
have to be distinguished from the data obtained for $L=2p$.
Indeed, in this parameter regime, the spin 
correlation length along the chains becomes of the order of the system size
so that ladders with $L=2p$ and 2 holes suffer from a small
magnetic frustration induced by the boundary conditions.
An accurate finite size scaling analysis can nevertheless be realized by 
considering the data for $L=2p+1$ which show again a very systematic
behavior as a function of $L$; in fact, although
the behavior of the data sets corresponding to ladders with 
$L=4p+1$ ($L=4p+3$) 
rungs and PBC follow only roughly the same trend as the data set 
corresponding to ladders with $L=4p+3$ ($L=4p+1$) rungs and ABC, 
averaging over PBC and ABC leads to a single remarkably smooth behavior of 
$\Delta_B$ vs $L$ as observed in Fig.~\ref{Delta_vs_L}(b).
A similar procedure can be followed for the data obtained with $L=2p$
as shown also in Fig.~\ref{Delta_vs_L}(b) but this extrapolation is 
probably less reliable for the reasons stated above. 
In all cases, the $L\rightarrow\infty$ extrapolation is performed 
according to $\Delta_B(L)=\Delta_B^\infty +A\frac{1}{L}\exp{(-L/\xi)}$, where 
$\Delta_B^\infty$, $\xi$ and $A$ are free parameters determined
from a fit to the data. 

\begin{figure}[htbp]
\begin{center}
\vspace{-0.8truecm}
\psfig{figure=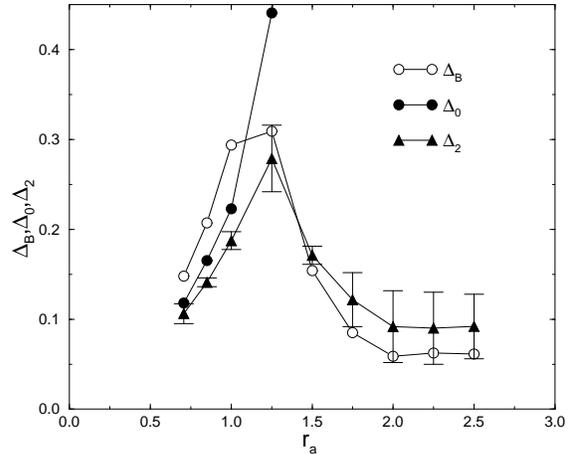,width=8.6truecm,angle=-90}
\end{center}
\caption{Extrapolated two hole binding energy (open symbols) as a function of
the anisotropy $r_a=t_\perp/t_\parallel$ for $J_\parallel=0.5$. 
Spin gaps in the half-filled and two hole doped GS
are also shown for comparison (filled symbols).
}\label{Delta_vs_tperp}
\end{figure}

The extrapolated values of $\Delta_B$ are displayed in 
Fig.~\ref{Delta_vs_tperp} as a function of $t_\perp$ for 
$J_\parallel=0.5$. A positive binding energy (implying the 
stability of the hole pair) is obtained for all parameters considered here.
Recent DMRG work using
clusters with up to $2 \times 30$ sites and two holes have
calculated the binding energy in the isotropic limit~\cite{gazza}.
The result $\Delta_B \sim 0.2$ reported there is very close to 
the number found by our size extrapolation in Fig.~\ref{Delta_vs_tperp}, 
giving extra support to our procedure.
For comparison, the spin gaps $\Delta_0$ and $\Delta_2$
in the undoped GS and 2 hole doped GS, respectively are also shown. 
The scaling behavior of 
$\Delta_2$ is very similar to that of $\Delta_B$ as can be seen in 
Figs.~\ref{Delta_vs_L}(c,d) (although finite size effects are larger)
and  the same procedures to obtain 
the extrapolations to $L=\infty$ have been used.
Our extrapolations for $\Delta_0$ are in good agreement with previous
ED~\cite{SG_ED} and QMC~\cite{SG_QMC} estimates or with the weak coupling 
limit behavior $\Delta_0\sim 0.41 J_\perp$.
The behaviors of $\Delta_B$ and $\Delta_2$ with $t_\perp$ 
are very similar, and both have  a pronounced maximum of $\Delta_B$ 
at $t_\perp/t\sim 1.25$. In fact it is expected that in the 2 hole doped GS
the lowest $S=1$ excitations can be obtained by either (i) breaking up a hole
pair and flipping one of the unpaired spins or (ii) making 
a spin excitation away
from the hole pair (which is supposed to have a finite size).
Therefore, one expects $\Delta_2 ={\rm Min}\, \{ \Delta_B,\Delta_0 \}$.
Our data indeed suggest two regimes: (i) for $t_\perp/t_\parallel \leq 1.25$,
the binding energy of a pair is larger than the spin gap in the undoped system.
This extra stability of the hole pair
is probably due to the strong antiferromagnetic correlations
within each chain. Note that, in this regime, there are some small 
discrepancies between $\Delta_2$ and $\Delta_0$. This effect is probably due
to the fact that the size of the pair is particularly large for such 
parameters and the two regions corresponding to the undoped spin liquid 
and the hole pair cannot truly separate in the clusters that can be
handled computationally
leading to strong finite size effects for $\Delta_2$. 
(ii) for $t_\perp/t_\parallel \geq 1.25$,
the hole binding energy strongly decreases and the lowest spin excitation
is obtained by breaking a hole pair i.e. $\Delta_2\sim \Delta_B$. 
Note that for even larger ratios $t_\perp/t_\parallel$ (typically 
$t_\perp/t_\parallel > 2.5$), the binding energy increases again. 
Clearly, this is an artificial effect due to the fact that, in our model,
the rung magnetic coupling scales like $t_\perp^2$ and becomes unphysically 
large compared to $t_\perp$ for large enough $t_\perp$. 
In that case, $\Delta_B\simeq J_\perp -2t_\perp$
which approaches the spin gap $\Delta_0$ for large $t_\perp$. 

It is interesting to compare the results of Fig.~\ref{Delta_vs_tperp}
with the previous study of the collective modes of the $t-J$ 
ladder~\cite{modes_ladder}. On general grounds, two 
collective spin modes of momenta $q_\perp=0$
and $q_\perp=\pi$ are expected in a doped spin ladder.
Both modes are gapped at moderate doping~\cite{modes_ladder}. From 
a careful examination of the quantum numbers of the various 
spin excitations shown in Fig.~\ref{Delta_vs_tperp}, one can safely
study, at vanishing doping (i.e.
for 2 holes in an infinitely large system), each low energy excitation. 
The collective $q_\perp=\pi$ 
spin mode corresponds to the spin excitation of energy $\Delta_0$ 
characteristic of the undoped system (crudely an excitation of a singlet rung
into a triplet). 
On the other hand, the $q_\perp=0$ spin 
mode is associated to the breaking of a hole 
pair of energy $\Delta_B$. From our previous analysis of the data, a level crossing 
occurs between these two types of excitations around $t_\perp\simeq 1.25$ 
producing a cusp-like maximum of $\Delta_2$. 

Materials corresponding to the regime $t_\perp/t_\parallel> 1.25$ 
should be particularly interesting to be studied by 
Inelastic Neutron Scattering (INS) experiments at small
doping. Indeed, the above calculation predicts that, under light doping,
spectral weight in the dynamical spin structure factor $S({\bf q},\omega)$ 
should appear {\it within} the spin gap of the undoped material. 
This new $q_\perp=0$ magnetic structure whose total weight 
should roughly scale
with the doping fraction corresponds to the excitations of hole pairs into
two separate holes in a triplet state. The corresponding energy scale for
such excitations can 
be much lower than the spin gap of the undoped spin liquid GS ($q_\perp=\pi$).

The maximum observed in $\Delta_B$ for the $t-J$ model at $J_\parallel=0.5$
as a function of $t_\perp$ has similarities with the behavior 
of the pair-pair correlation obtained in the Hubbard model 
at very small hole doping~\cite{Noack} $n=0.9375$ which also shows a maximum
(around $t_\perp\simeq 1.4$ for $U=8$). In Ref.~\cite{Noack}, 
this particular value of $t_\perp$ was associated with the situation 
where the chemical potential
coincides almost exactly with the top of the lower bonding band and
with the bottom of the upper antibonding band. 
In that case, one expect a particularly large density of state at the
chemical potential (see also Ref.~\cite{naza}).
However, such a correspondence was made possible at smaller $U$ only
(due to difficulties to obtain accurate QMC calculations of 
dynamical quantities at intermediate and large values of $U$).
The spectral function $A({\bf q},\omega)$ shown
in Fig.~\ref{Akw1_tJ}(b) was obtained in the two hole GS of the $2\times 8$ ladder
for a choice of parameters ($J_\parallel=0.4$ and $t_\perp=1$) close to the 
ones producing the maximum of $\Delta_B$ in Fig.~\ref{Delta_vs_tperp}.
Fig.~\ref{Akw1_tJ}(b) clearly shows a large density of states in the vicinity
of the chemical potential due to the flatness of the dispersion around
${\bf q}=(0,\pi)$ or ${\bf q}=(\pi,0)$. 
This situation corresponds to the cross-over between the two band and four
band insulator regimes observed at half-filling~\cite{Hubbard_neq1}.
It is also interesting to note that a small depression of the density of state 
is visible in Fig.~\ref{Akw1_tJ}(b) at the chemical potential.
This could be interpreted as a small gap associated to the existence of
a bound pair. More generally, in the so called C1S0 
phase~\cite{phase_diagram,weak_coupling,mueller} where the spin gap survives,
one expects to see its signature in $A({\bf q},\omega)$ as a gap at the 
chemical potential. However, the energy scale of the spin gap 
is small (see e.g. the order of magnitude of
$\Delta_2$ in Fig.~\ref{Delta_vs_tperp}) compared to the various features 
that appear in 
$A({\bf q},\omega)$ and thus, in most cases, its manifestation in 
$A({\bf q},\omega)$ cannot be observed on small lattices.  
In the recent studies using a reduced
Hilbert space, the observation of a gap caused by pairing in the
spectral function required the use of clusters with $2 \times 16$
and $2 \times 20$ sites~\cite{new2}.

\subsection{Pair-pair correlations}

ED studies supplemented by conformal invariance arguments 
suggest that in the doped spin gap
phase (C1S0) of the isotropic $t-J$ ladder (where pairs are formed 
according to the previous analysis) algebraic superconducting and 
$4k_F$-CDW correlations are competing~\cite{phase_diagram}.
At small $J/t$ ratio, the CDW correlations dominate while above 
a moderate critical value of $J/t$ coherent hopping of the pairs takes over.
The aim of the present Subsection is to investigate the role of the anisotropy
$t_\perp/t_\parallel$ by a direct calculation of the pair-pair correlation 
as a function of distance.
As previously, in the case 
of the $t-J$ model,
a rung magnetic coupling $J_\perp=J_\parallel(t_\perp/t_\parallel)^2$ 
is used.

Superconducting correlations can be evidenced from a study of the long 
distance behavior of the pair hopping correlation,
\begin{equation}
C_{S}(r-r')=\big< \Delta^\dagger(r)\Delta(r')\big>\, ,
\label{pairing}
\end{equation}
where $\Delta^\dagger(r)$ is a creation operator of a pair centered
at position labeled by $r$. 
Although the best choice of $\Delta^\dagger(r)$ clearly depends on the 
internal structure of the hole pair~\cite{sqrt2} as discussed later,
it should exhibit general symmetry properties associated to the
quantum numbers of the hole pair found in the previous Subsection:
(i) $\Delta^\dagger(r)$ is a singlet operator and (ii) it is even with
respect to the two reflection symmetries along and perpendicular to
the ladder direction (and centered at position $r$).
The static correlation function of Eq.~\ref{pairing} can be interpreted as
a coherent hopping of a pair centered at position $r$ to a new position
$r'$. 

According to conformal invariance, in a strictly 1D ladder
(which is the case studied here) the pair hopping correlation 
exhibits a power-law behavior at large distances $|r-r'|$,
\begin{equation}
C_{S}(r-r')\sim 1/|r-r'|^\frac{1}{2K_\rho}\, ,
\end{equation}
where the exponent $K_\rho$ was calculated in the weak coupling 
limit~\cite{weak_coupling} or in the isotropic $t-J$ ladder by 
ED methods using conformal invariance relations~\cite{phase_diagram}.  
Superconducting correlations dominate when $K_\rho>1/2$ which occurs
for $J/t>0.3$ in the lightly doped isotropic $t-J$ ladder~\cite{phase_diagram}.
Using a DMRG approach, the behavior of $C_{S}(r-r')$ with the usual BCS
bond pair operator,
\begin{equation}
\Delta(i) = c_{i,1;\uparrow} c_{i,2;\downarrow}
- c_{i,1;\downarrow} c_{i,2;\uparrow}   
\, ,
\label{bcs}
\end{equation}
can also be obtained directly, leading, in the
case of the isotropic $t-J$ model~\cite{Hayward95}, to a good agreement 
with the ED results. 
More recently, this study was extended to the anisotropic Hubbard 
ladder~\cite{Noack} showing a pronounced peak of the long-distance pair 
correlations as a function of $t_\perp$. 

Here, as a complementary study of the analysis presented for 
the binding energy in
the previous Subsection, the behavior of the pair correlation
function of the BCS-like operator of Eq.(~\ref{bcs}) is compared against
the case where 
a spatially-extended pair operator is used. The first motivation to
introduce this new pair operator is due to the structure of the hole pair; 
indeed, it turns out that configurations in which the two holes sit 
along the diagonal of a plaquette carry a particularly large 
weight in the 2-hole GS both in the case of the 
2D $t-J$ model~\cite{sqrt2} or in the case of the $t-J$ 
ladder~\cite{White}. This feature seems counterintuitive 
in a two-hole bound state of $d_{x^2-y^2}$ character, as it is
the case  e.g. in 2D (for ladders, this symmetry is only approximate),
since the pair state is odd with respect to a reflection along the plaquette
diagonals.
However, it has been observed~\cite{retardation} that retardation 
provides in fact a simple physical explanation of this apparent paradox.
Secondly, it is clear that pairs extending into a larger region of space 
can acquire more internal kinetic energy and they are less sensitive 
to short distance electrostatic repulsion. 

\begin{figure}[htbp]
\begin{center}
\vspace{-1.05truecm}
\psfig{figure=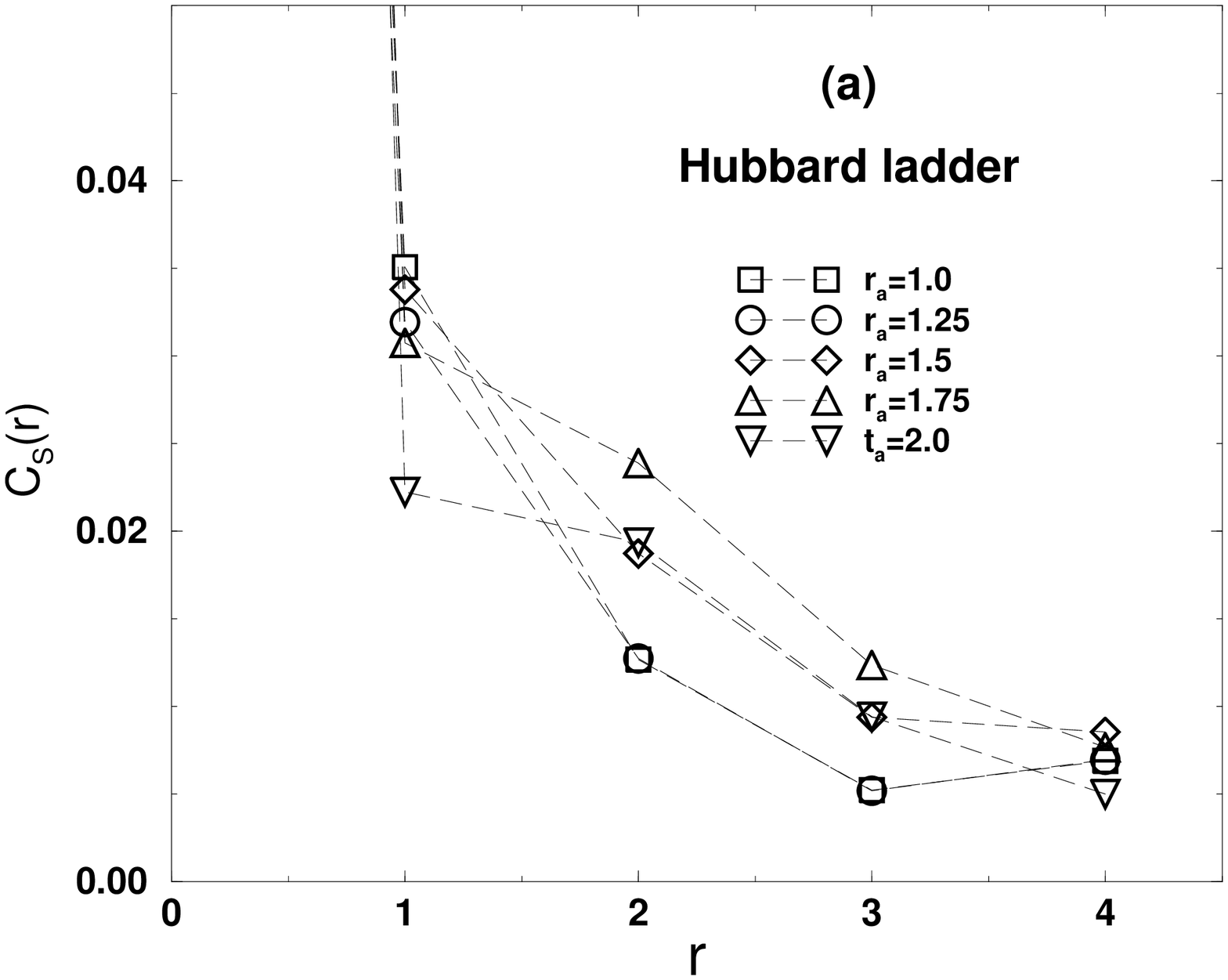,width=8.6truecm,angle=-90}
\vspace{-1.05truecm}
\psfig{figure=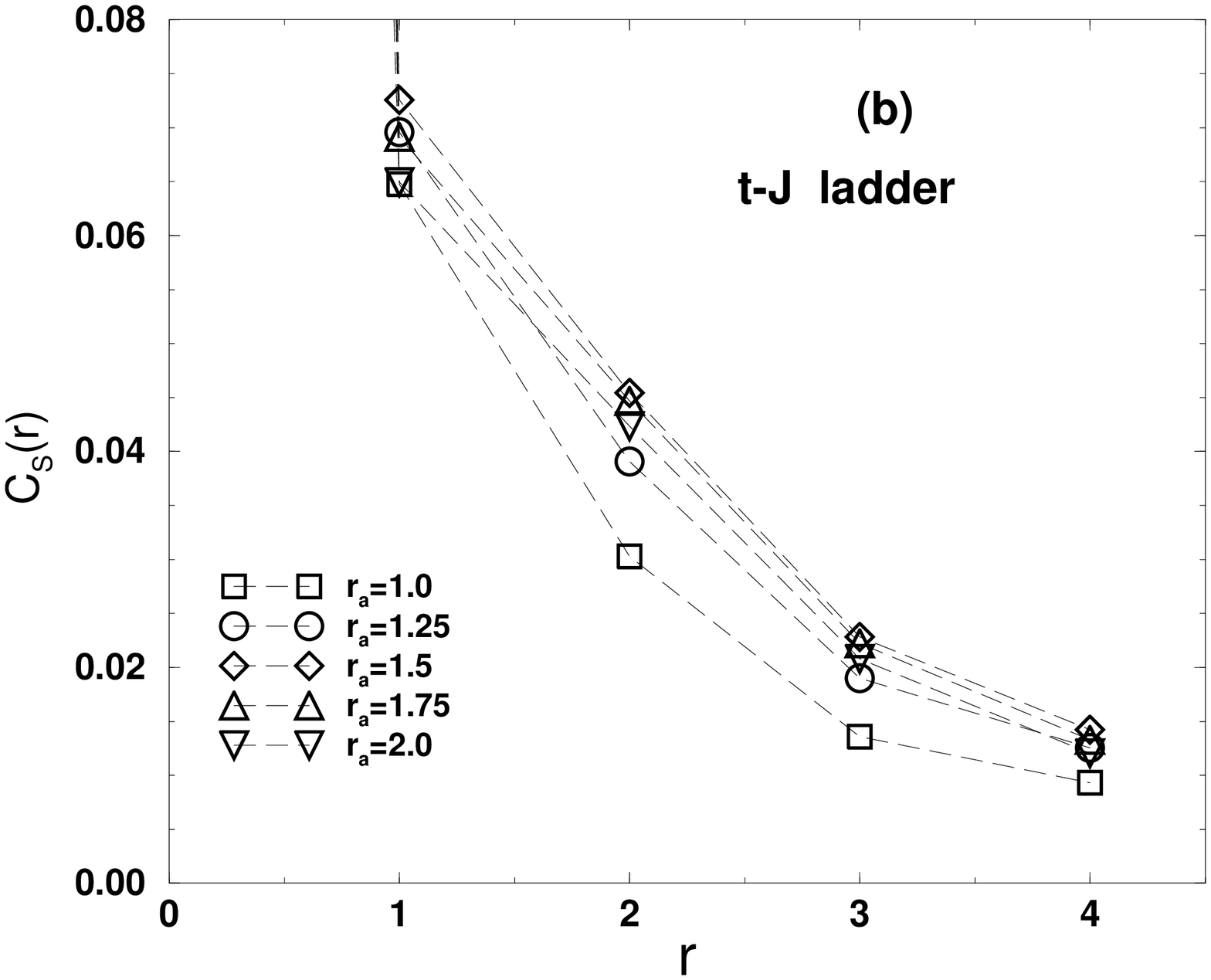,width=8.6truecm,angle=-90}
\vspace{-1.05truecm}
\psfig{figure=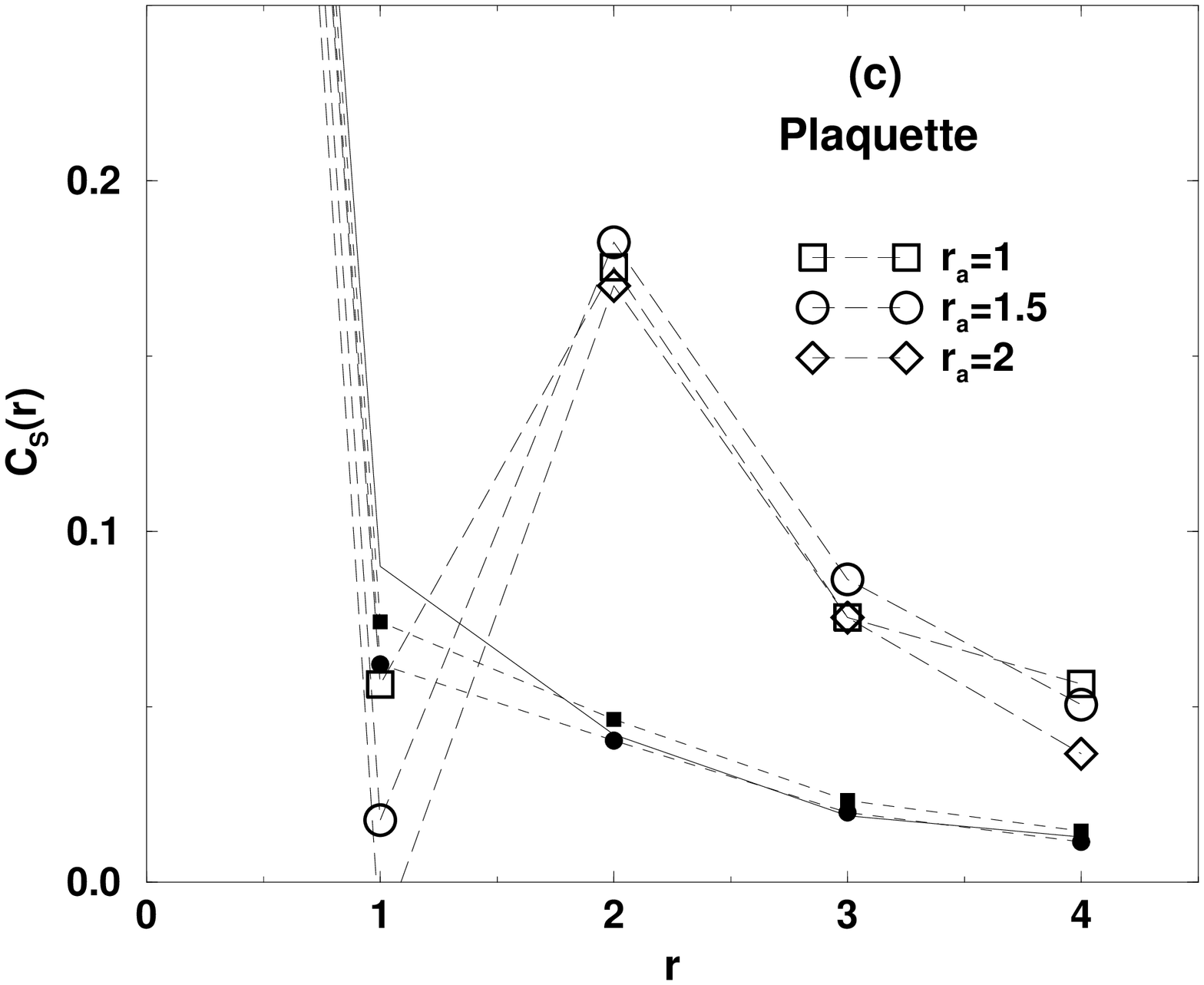,width=8.6truecm,angle=-90}
\end{center}
\caption{
Pair-pair correlation function vs distance calculated on $2\times 8$ clusters
at density $n=0.75$ with PBC in the chain direction. The values of the 
anisotropy $r_a=t_\perp/t_\parallel$ are indicated on the plot.  
(a) rung-rung correlations in the Hubbard ladder for $U=10$;
(b) rung-rung correlations in the $t-J$ ladder for $J_\parallel=0.4$;
(c) plaquette-plaquette correlations (open symbols) in the 
$t-J$ ladder for $J_\parallel=0.4$. For comparison, some of the 
correlations of the rung pair operator of (b) are also reproduced
(small full symbols) on the same plot.
}
\label{pair_pair}
\end{figure}

To study the influence of the spatial extension of the 
pair operator $\Delta(r)$ on the pair-pair correlation, 
following Ref.~\cite{sqrt2} here a {\it plaquette} pair
operator
is defined as
\begin{eqnarray}
\Delta(i+1/2)&=&
({\bf S}_{i,2}-{\bf S}_{i+1,1})\cdot 
{\bf T}_{i,1;i+1,2}  \nonumber \\
&-&({\bf S}_{i,1}-{\bf S}_{i+1,2})\cdot 
{\bf T}_{i+1,1;i,2} \, 
\label {nnnpairing}
\end{eqnarray}
\noindent  
where ${\bf T}_{i,\alpha;j,\beta}=\frac{1}{i}c_{i,\alpha;\sigma}
(\sigma_y {\vec\sigma})_{\sigma\sigma^\prime}c_{j,\beta;\sigma^\prime}$ is the
regular (oriented) spin triplet pair operator~\cite{note_plaquette}.
Physically, $\Delta^\dagger(i+1/2)$ creates a singlet pair centered on 
a plaquette in a $d_{x^2-y^2}$ state with holes located
along the diagonals of the plaquette (at distance $\sqrt{2}$). 
The interpretation of this operator is simple: starting from a 
hole pair located on a rung, the hopping of one of the holes 
by one site along the leg-ladder leaves behind a spin with the opposite
orientation than the local AF pattern. This argument naturally 
leads to a 3-body problem~\cite{retardation} involving a triplet hole pair and
a local spin flip (of triplet character). 
Formally, this picture is equivalent to introducing some retardation 
in the usual BCS operator i.e.  the two holes can be created 
at two different times separated by an amount $\tau$ e.g.
by applying $c_{i,1;\uparrow}(\tau/2) c_{i,2;\downarrow}(-\tau/2)$ on the
AF background. The expansion of this new operator to order $\tau^2$
then leads to the various terms of Eq.(~\ref{nnnpairing}).
Alternatively, $\Delta^\dagger(i+1/2)$ can also be viewed as the simplest
$d_{x^2-y^2}$ operator of global singlet character creating
a pair on the diagonals of a plaquette. This result can be deduced from
simple symmetry considerations~\cite{sqrt2}. 

Our results for $C_s(r)$ in the case of the rung BCS-like operator are 
shown in Fig.~\ref{pair_pair}(a) and (b) for the Hubbard and $t-J$ ladders,
respectively. Both sets of data are consistent with the power law decay
and show a clear increase of the correlations at intermediate distances. 
In the case of the $t-J$ ladder at $n=0.75$, the maximum occurs for
$t_\perp\simeq 1.5$, a value slightly larger than the characteristic
value corresponding to the maximum of $\Delta_B$. 
According to Figs.~\ref{Akw1_Hubbard} and \ref{Akw2_tJ} showing the
single particle spectral functions for almost identical parameters, 
this specific value of $t_\perp$ seems to correspond to the case where
the chemical potential sits in the vicinity of a maximum of the density of 
states generated by very flat bands at the band edge (as suggested in 
Ref.~\cite{Noack} and in agreement with the general 
ideas discussed in Ref.~\cite{naza}).
On the other hand, it is likely that the maximum of $\Delta_B$ does not
occur at exactly the same value of $t_\perp$ but rather at a somewhat 
smaller value. 

The plaquette pair-pair correlations are shown in Fig.~\ref{pair_pair}(c).
At short distance $r=1$, the correlations are suppressed reflecting
the spatial extension of the pair operator. 
At larger distances, $r\ge 2$, a significant overall increase is 
observed compared to the case of the rung
operator, showing that indeed the use of ``extended'' operators to
capture
the usually weak signals of superconductivity in doped antiferromagnetic
systems is a promising strategy~\cite{bob}. Note that, apart from this overall
factor, the functional form of the decay seems to be identical to
the one obtained for the rung operator (as can be checked quantitatively).

\section{Conclusions}

In this paper dynamical properties of anisotropic ladders have been
investigated using the one-band Hubbard and $t-J$ models. An analysis
based
on the local-rung approximation explains a considerable part of the
numerical results. 
In particular, the existence of a metal-insulator transition at
quarter filling which can be justified in such an analysis was indeed 
numerically seen for increasing anisotropy ratio.
Flat quasiparticle dispersions at the chemical
potential are observed in regions of parameter space where pairing
correlations are robust. A finite-size scaling of the binding energy and
the spin-gap show that these quantities change with the anisotropy ratio
in a manner similar as the pair correlations do. In agreement with
previous results, it is observed that superconducting correlations are
maximized for  anisotropic systems, with couplings along rungs 
slightly larger than along the legs.

\section{acknowledgments}

E.~D. is supported by the 
NSF grant DMR-9520776. D.~P. and J.~R. thank IDRIS, Orsay (France) for
allocation of CPU time on the C94, C98 and T3E Cray supercomputers. 
J.~R. acknowledges partial support from the Ministry of Education (France) and 
the Centre National de la Recherche Scientifique (CNRS).


\begin{thebibliography}{}


\bibitem{Review_ladder} For a review see e.g.
E.~Dagotto and T.M. Rice,
{\em Science}, {\bf 271}, 618 (1996).

\bibitem{Uehara96}
M.~Uehara, T.~Nagata, J.~Akimitsu, H.~Takahashi, N.~M\^ori, and K.~Kinoshita,
{\em J. Phys. Soc. Japan.}, {\bf 65}, 2764 (1996).

\bibitem{Millet} P. Millet et al., Phys. Rev. B {\bf 57}, xxx (1998). 

\bibitem{xrays_NaV2O5} H. Smolinski, C. Gros, W. Weber, U. Peuchert, G. Roth,
M. Weiden and C. Geibel, cond-mat/9801276 (1998).

\bibitem{Dagotto92}
E.~Dagotto, J.~Riera and D.J.~Scalapino,
\newblock {\em Phys. Rev. B}, {\bf 45}, 5744 (1992); see also
H. J. Schulz, \newblock {\em Phys. Rev. B}, {\bf 34}, 6372 (1986);
E.~Dagotto and A.~Moreo,
\newblock {\em Phys. Rev. B}, {\bf 38}, 5087 (1988).

\bibitem{Azuma94}
M.~Azuma, Z.~Hiroi, M.~Takano, K.~Ishida, and Y.~Kitaoka,
{\em Phys. Rev. Lett.}, {\bf 73}, 3463 (1994).


\bibitem{Hiroi95}
Z.~Hiroi and M.~Takano, {\em Nature}, {\bf 377}, 41 (1995).

\bibitem{pairing}
M.~Sigrist, T.M. Rice, and F.C. Zhang, {\em Phys. Rev. B},  {\bf 49},12058
(1994);
H.~Tsunetsugu, M.~Troyer, and T.M. Rice, {\em Phys. Rev. B}, {\bf 49},16078
(1994).

\bibitem{Jerome} A. Mayaffre et al., {\em Science}, {\bf 279}, 345 (1998).

\bibitem{modes_ladder}
D.~Poilblanc, D.~J. Scalapino, and W.~Hanke, {\em Phys. Rev. B}, {\bf 52},
6796 (1995). 

\bibitem{weak_coupling}
L.~Balents and M.P.A.~Fisher,
{\em Phys. Rev. B}, {\bf 53} 12133 (1996);
H.J. Schulz, {\em Phys. Rev. B}, {\bf 54} R2959 (1996).

\bibitem{Hayward95}
C.~Hayward, D.~Poilblanc, R.M.~Noack, D.J.~Scalapino, and W.~Hanke, {\em
Phys. Rev. Lett.}, {\bf 75}, 926 (1995).

\bibitem{phase_diagram}
C.~Hayward and D.~Poilblanc, {\em Phys. Rev. B}, {\bf 53}, 11721 (1996).
See also
H.~Tsunetsugu, M.~Troyer, and T.M. Rice, {\em Phys. Rev. B}, {\bf 51},
16456 (1995).

\bibitem{noteCnSm} To distinguish the different phases, the notation
C$n$S$m$ was introduced in Ref.~\protect\cite{weak_coupling}
to label a phase with $n$ gapless charge modes and $m$ gapless
spin modes.

\bibitem{Orignac} E.~Orignac and T.~Giamarchi, {\em Phys. Rev. B} {\bf 56},
7167 (1997). 
 
\bibitem{mueller} T.F.~M\"uller and T.M.~Rice, cond-mat/9802297 
preprint (1998).

\bibitem{Eccleston} R.S.~Eccleston, M.~Uehara, J.~Akimitsu, H.~Eisaki, 
N.~Motoyama and S. Uchida,  cond-mat/9711053 (1997).

\bibitem{Johnston} D.C.~Johnston, {\em Phys. Rev. B}, {\bf 54}, 13009 (1996). 

\bibitem{Horsch} P. Horsch and F. Mack, cond-mat/9801316 (1998).

\bibitem{Augier1} D. Augier, D. Poilblanc, S. Haas, A. Delia and 
E. Dagotto, Phys. Rev. B {\bf 56}, R5732 (1997).

\bibitem{SG_anisotropy}
T.~Barnes, E.~Dagotto, J.~Riera and E.~Swanson,
{\em Phys. Rev. B}, {\bf 47}, 3196 (1993); see also S. Gopolan,
T. M. Rice and M.~Sigrist, {\em Phys. Rev. B}, {\bf 49}, 8901 (1994).

\bibitem{Hubbard_neq1} H. Endres, R. M. Noack, W. Hanke, D. Poilblanc
and D. J. Scalapino, {\em Phys. Rev. B} {\bf 53}, 5530 (1996).

\bibitem{Maekawa} C. Kim, A. Y. Matsuura, Z.-X. Shen, N. Motoyama, H. Eisaki,
S. Uchida, T. Tohyama and S. Maekawa, {\em Phys. Rev. Lett.} {\bf 77},
4054 (1996); C. Kim {\it et al.}, {\em Phys. Rev. B} {\bf 56}, 15589 (1997).

\bibitem{Noack} R.M.~Noack, N.~Bulut,
D.~J.~Scalapino and M.~G.~Zacher, {\em Phys. Rev. B} {\bf 56},
7162 (1997). 

\bibitem{new1} E. Dagotto, G. Martins, J. Riera and A. Malvezzi,
preprint. 

\bibitem{new2} G. Martins, J. Riera, and E. Dagotto, unpublished.

\bibitem{note1} In this case, one cannot completely 
exclude an exponentially small single particle gap.

\bibitem{Haas} Calculations of $A({\bf q},\omega)$ in the {\it isotropic}
$t-J$ ladder can be found e.g. in S. Haas 
and E. Dagotto, Phys. Rev. B {\bf 54}, R3718 (1996). 

\bibitem{White} S.~White and D.J.~Scalapino, {\em Phys. Rev. B} {\bf 55},
6504 (1997). 

\bibitem{gazza} C. Gazza et al., preprint (cond-mat/9803314).

\bibitem{SG_ED} M.~Reigrotzki, H.~Tsunetsugu and T.M.~Rice,
{\em J. Phys. C} {\bf 6}, 9235 (1994).

\bibitem{SG_QMC} M.~Greven, R.J.~Birgeneau and U.-J.~Wiese,
{\em Phys. Rev. Lett.} {\bf 77}, 1865 (1996).

\bibitem{sqrt2} D. Poilblanc, {\em Phys. Rev. B} {\bf 49}, 1477 (1994).

\bibitem{retardation} J.~Riera and E.~Dagotto,  {\em Phys. Rev. B} {\bf 57}, 
xxxx (1998).

\bibitem{note_plaquette} D.P. thanks D.J.~Scalapino for pointing out
a misprint in note [21] of Ref.~\protect\cite{sqrt2}.
The correct expression for ${\bf S}_{k}\cdot 
{\bf T}_{i; j}$ reads $S_{ k}^Z (c_{i;\uparrow}
c_{j;\downarrow}-c_{j;\uparrow}c_{i;\downarrow})
-S_{ k}^+ c_{ i;\uparrow}c_{ j;\uparrow}
+S_{ k}^- c_{ i;\downarrow}c_{ j;\downarrow}$.

\bibitem{naza} 
E. Dagotto, A. Nazarenko and A. Moreo,
Phys. Rev. Lett. {\bf 74}, 310 (1995).

\bibitem{bob} Related ideas where presented years ago
in the same context by 
E. Dagotto, and J.R. Schrieffer, Phys. Rev. {\bf B
43}, 8705 (1991).

\end{thebibliography}
\end{document}